\documentclass[apj]{emulateapj}
\usepackage{apjfonts}
 
\slugcomment{Submitted to {\it{The Astrophysical Journal}}}
\shortauthors{Gonzalez et al.}
\shorttitle{Intracluster Light:  Relationship to BCG Halos}

\newcommand \sbu {mag arcsec$^{-2}$}
\newcommand \ho  {}
\newcommand \ddev {2-deV }
\newcommand \dev{$r^{1/4}$ }
\newcommand \ser {S\'ersic }

\begin{document}
\title{ Intracluster Light in Nearby Galaxy Clusters: Relationship to the
Halos of Brightest Cluster Galaxies }
  
\author{Anthony H. Gonzalez\altaffilmark{1}}
\affil{Department of Astronomy, University of Florida, Gainesville, FL 32611}
\email{anthony@astro.ufl.edu}
\and
\author{Ann I. Zabludoff and Dennis Zaritsky}
\affil{Steward Observatory, University of Arizona, 933 North Cherry Avenue, Tucson, AZ 85721}
   
\altaffiltext{1}{NSF Astronomy and Astrophysics Postdoctoral Fellow}
    
\begin{abstract}                                                   
We present a detailed analysis of the surface brightness distribution of the
brightest cluster galaxy (BCG) in each of 24 galaxy clusters at
$0.03<z<0.13$. We use two-dimensional profile fitting to model the surface
brightness out to $r=300$ \ho kpc for each BCG, comparing $r^{1/4}$,
$r^{1/n}$, and double $r^{1/4}$ models.  We obtain statistically superior fits
using a two component model consisting of a pair of $r^{1/4}$ profiles with
independent scale lengths, ellipticities, and orientations. The two component
model can simply reproduce the observed position angle and ellipticity
gradients, which cannot generally be explained purely by triaxiality.  The
inner component of our two component model has properties similar to a typical
massive elliptical galaxy and is clearly associated with the BCG.  The outer
component is 10-40 times larger in scale, has $\sim10$ times the total
luminosity of the inner component, and exhibits a steeper $<\mu>-r_e$ relation
than that of the elliptical fundamental plane.  We interpret this outer
component as a population of intracluster stars tracing the cluster potential.
The two components are strongly aligned ($|\Delta\theta|<10\degr$) in roughly
40\% of the clusters.  When they are not aligned, the components tend toward
high degrees of misalignment, suggesting that accretion of infalling material
may change the orientation of some BCGs for a time.  The extent of the outer
component and its similar elongation to published cluster galaxy distributions
indicates that the evolution of the intracluster light is tied to the cluster
as a whole rather than to the BCG.
\end{abstract}


\section{Introduction}
\label{sec:intro}

One of the most distinctive characteristics of brightest cluster galaxies
(BCGs) is that many exhibit shallower extended surface brightness profiles
than their lower mass elliptical counterparts. This trait was first recognized
by \citet{mms1964}, who observed that some BCGs emit more light at large radii
than would be expected for an \dev profile. This excess emission, which is
often lumped into the term ``cD" envelope,\footnote{This emission is
differentiated into categories by some experts (see \citealt{schombert1986}).}
was confirmed by \citet{oemler73,oemler76} and
\citet{schombert1986,schombert1987,schombert1988} in a subset of BCGs.

What is the nature of this excess luminosity at large radii?  Schombert, who
considered two-component models when fitting surface brightness profiles to
$r\sim200-1500$ \ho kpc, argued that the extended light is physically distinct
from the central galaxy.  In this picture, the excess light component is
thought to arise from a population of intracluster stars produced by the tidal
stripping and disruption of cluster galaxies
\citep[c.f.][]{gallagher1972,dressler1979,richstone1983,miller1983,
malamuth1984,merritt1984,larsen2004,murante2004,willman2004}.  Simulations
predict that this intracluster light (ICL) is comprised of old stars that are
dynamically distinct from the stars in the BCG \citep{larsen2004,murante2004}
and that production of intracluster stars is an ongoing process
\citep{willman2004}.  If correct, this model implies that the properties of
the ICL, including its fractional contribution to the total cluster luminosity
and its spatial distribution, are sensitive probes of the physical processes
driving cluster formation and the evolution of cluster galaxies.

Subsequent studies using CCD detectors have generally lacked either the
statistics or the radial coverage required to test this
interpretation. Moreover, most recent measurements of individual BCG profiles
out to large radii ($r\ga200$ \ho kpc) have failed to detect a second
component.  With the exception of \citet{feldmeier2002}, these studies find
profiles that are roughly consistent with a pure \dev law
\citep{uson1990,uson1991,scheick1994,gonzalez2000}.  With this program we aim
to utilize a large sample of BCGs to constrain the ubiquity of the outer
component and to quantify the properties of this component if it exists.

Our work extends beyond previous studies in several ways. First, by utilizing
highly uniform drift-scan data we provide the first observations that recover
the profiles for a statistical sample of BCGs to large radii using modern CCD
photometry. Specifically, we present the surface brightness profiles to
$r\ga300$ \ho kpc for 24 BCGs.  Second, scanning provides long, contiguous
images obtained with the same CCD. The importance of a precise and accurate
sky level is paramount in this type of study, and the geometry of drift-scans
facilitates this endeavor.  Third, while previous studies typically either
azimuthally average or model the surface brightness profile along the
semi-major axis, we employ full two-dimensional modelling. This approach
enables us to model ellipticity or position angle gradients that may be
present and to simultaneously model multiple objects. The latter minimizes the
impact on the profile of other galaxies projected near the center of the BCG.

The organization of this paper is as follows. In \S\ref{sec:data} we discuss
the data and preliminary reductions. In \S\ref{sec:profiles} we describe the
modelling of the galaxies and present the resulting profiles. In
\S\ref{sec:results} we find that two-component fits provide better fits to the
profiles than one-component models, compare the relative properties
(luminosity, effective radius, ellipticity, position angle, centroid) of the
two components, discuss the fundamental plane for both components, and
consider how one-component fits impact determination of the BCG total
magnitudes and fundamental plane.  We assess the robustness of the derived
structural parameters in the Appendix.  Finally, we summarize our conclusions
in \S\ref{sec:discussion}. Throughout the paper we assume a cosmology with
$H_0=70$ km s$^{-1}$ Mpc$^{-1}$, $\Omega=0.3$, and $\Lambda=0.7$.

Our results provide important tests for the new generation of cosmological
cluster simulations that have both the force resolution and complexity to
generate intracluster stars \citep{larsen2004,murante2004, willman2004} and
that are beginning to make predictions for the properties of this population.

\section{Data and Reductions}
\label{sec:data}

\subsection{Sample Definition}

We observed a total of 30 clusters spanning a range of velocity dispersions
and Bautz-Morgan types \citep{bautz1970}.  Our sample consists of nearby
clusters (published redshifts $0.08<z<0.12$) that contain BCGs with position
angles within 45 degrees of the east-west axis (our drift scan direction).  In
this paper we focus on the 24 clusters with a single, dominant BCG that show
no evidence of an ongoing merger.  We note that Bautz-Morgan type III clusters
are underrepresented in the full sample and that errors in the published
redshifts yield a wider actual redshift range ($0.03<z<0.13$).

\subsection{Data}
We obtained drift scan data during four separate runs on the Las Campanas 1m
Swope telescope between 1995 and 2000 (Table \ref{tab:obs}). Conditions were
photometric during all runs, with the data from 2000 having the darkest sky
and best seeing.  All imaging was obtained in Gunn $i$ using the Great Circle
Camera \citep{zaritsky1996}, and the photometry is calibrated to Cousins $I$
using Landolt standards. Individual drift scans are 2048$\times$10000 pixels
with a plate scale of 0.7$\arcsec$ pixel$^{-1}$, except for the scans of Abell
1651, which are 13000 pixels long.  Exposure times per scan vary depending
upon the declination of the target, ranging from 95-133 seconds, with seven
scans obtained for each target when possible.  Observational details for each
field, including total exposure time and seeing, are listed in Table
\ref{tab:obs}. For the 24 clusters that are the focus of this paper we also
include the Two Micron All Sky Survey (2MASS) Extended Source Catalog
designation for the brightest cluster galaxy.

\subsection{Reductions}
The two factors that typically limit photometric measurements of low surface
brightness features are variations in the detector response (i.e.~flatness)
and variations in the sky flux. Minimizing residual variations from both
sources is criticial for this program, as brightest cluster galaxies are
expected to have $\mu_I\approx26$ \sbu~at $r=300$ \ho kpc.  To reliably model
the profiles out to this radius, we require that this surface brightness
exceed the $3\sigma$ systematic uncertainty in the sky level.

\subsubsection{Flatfielding}
The basic reduction procedure is similar in form to that described by
\citet{gonzalez2000}.  With drift-scan imaging, pixel-to-pixel variation is
mitigated because data are clocked across the chip, and so sensitivity
variations are a concern only perpendicular to the readout direction (at a
level of $\sim2$\% in our raw data).  After subtracting the bias, we use the
scans to generate a one-dimensional flatfield, using sigma-clipped averaging
to compute the sensitivity for each column. The accuracy of this initial
flatfield is limited by both contamination from astronomical objects and
temporal sky variations along the direction of readout.

We generate a second-pass flatness correction by masking objects and modelling
the sky variations (details are given in the discussion of sky subtraction
below). This improved flatfield is then applied to the original
bias-subtracted images. We generate a single flatfield for each of the 1995
and 1999 observing runs, while for the 2000 run we have sufficient data to
divide the run into three two-night subsets and generate an individual
flatfield for each subset. The difference among these flatfields reflects the
remaining uncorrected sensitivity variations.  The rms difference between the
second and third subsets of the 2000 data is $0.05$\%, with no detectable
systematic variation.  The flatfield from the first two nights exhibits a
roughly linear systematic variation across the detector relative to the other
subsets at the level of $\pm0.17$\%, with an rms difference of $0.11$\%.  We
conclude that the residual flatness variations for a single scan in our data
are $\la0.20$\% (or $\mu_I\simeq 26.4$ \sbu), with column-to-column rms
variations $\simeq0.05$\% ($\mu_I\simeq 27.9$ \sbu).  Residual variations are
further mitigated by coadding the scans, and any remaining gradients (such as
the linear difference between flatfields noted above) are muted further by the
two-dimensional sky subtraction procedure described below.

\subsubsection{Subtraction of Sky Variations: First Pass}

The mean sky level typically changes by $\sim3$\% during 
single scan. To correct for this temporal variability, we first mask all
objects in a scan to prevent them from biasing our model of the variation.  We
create a binary mask using segmentation images output by SExtractor 2.3
\citep{bertin1996}, excluding all pixels within the detection and buffer
regions around each object.  For objects with an area in excess of 250
sq. arcsec (primarily saturated stars), the buffer region consists of pixels
within 150$\arcsec$ of the object detection region. For smaller objects the
buffer radius is either 50$\arcsec$ (for those objects with area $>$25
sq. arcsec) or 5$\arcsec$ (for those with area $<$25 sq. arcsec).  As an added
precaution, the entire region within $350\arcsec$ of the BCG is masked.
 
We then model sky variation along the scan with a five segment cubic spline.
Subtraction of this model provides a first-pass removal of the large-scale
temporal variations, which is a necessary prerequisite for co-adding.  A more
precise, two-dimensional modeling of the sky is performed at a later stage.

\begin{figure}
\epsscale{1.0}
\plotone{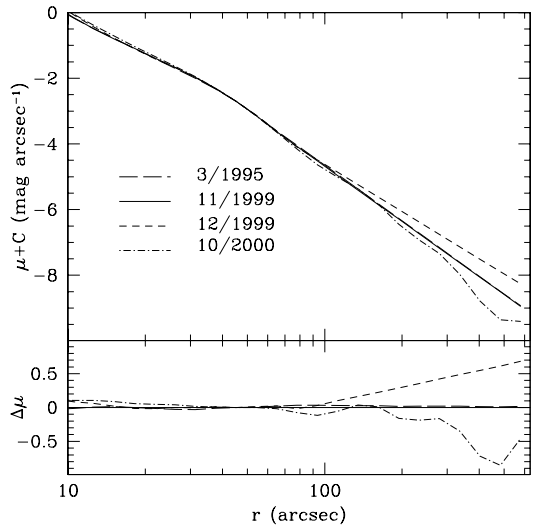}
\caption{$Top$: Empirical model PSFs of saturated stars generated for each
observing run, normalized to the same surface brightness at 50$\arcsec$.  The
PSF for the March 1995 run is the least well determined, being constructed
only from stars in the field of Abell 1651. We only plot $r>10\arcsec$, where
the profiles are independent of the mean seeing during the run.  $Bottom$:
Differences in the PSFs relative to the November 1999 run.  The PSFs for the
different runs agree to within roughly 30\% out to 200$\arcsec$.  }
\label{fig:psf}
\end{figure}

\subsubsection{PSF Modelling}
Following bias subtraction, flatfielding, and the first-pass sky subtraction,
we register and average the scans to generate a single image for each cluster.
A key concern at this stage is that the point spread function (PSF) of a
bright, saturated star can introduce appreciable gradients in the sky level
even at distances of several arcminutes. To minimize the impact of these stars
on our photometry, we construct an azimuthally averaged model of the PSF out
to $r=1000$ pixels ($700\arcsec$). A single model is generated for each
observing run via a luminosity weighted combination of all stars with
$m_I\la12.2$ in images from that run, masking nearby objects and the saturated
cores.  The resulting PSF models are shown in Figure \ref{fig:psf}, normalized
to have the same surface brightness at 50$\arcsec$ radius.  The PSF for the
1995 observing run is the least well constrained because it is generated using
only stars from the field of Abell 1651. Nonetheless, all profiles agree to
within $\sim30$\% out to 200$\arcsec$, and the difference between profiles at
$700\arcsec$ is not a dominant source of uncertainty in recovering the profile
of the BCG unless a star with $m_I\la7$ lies within this distance.  Abell 2376
and Abell 3809 are the only clusters in our sample with such a luminous nearby
star, while for Abell 2405 a bright star lies just outside this radius.

\begin{figure*}
\epsscale{0.48}
\plotone{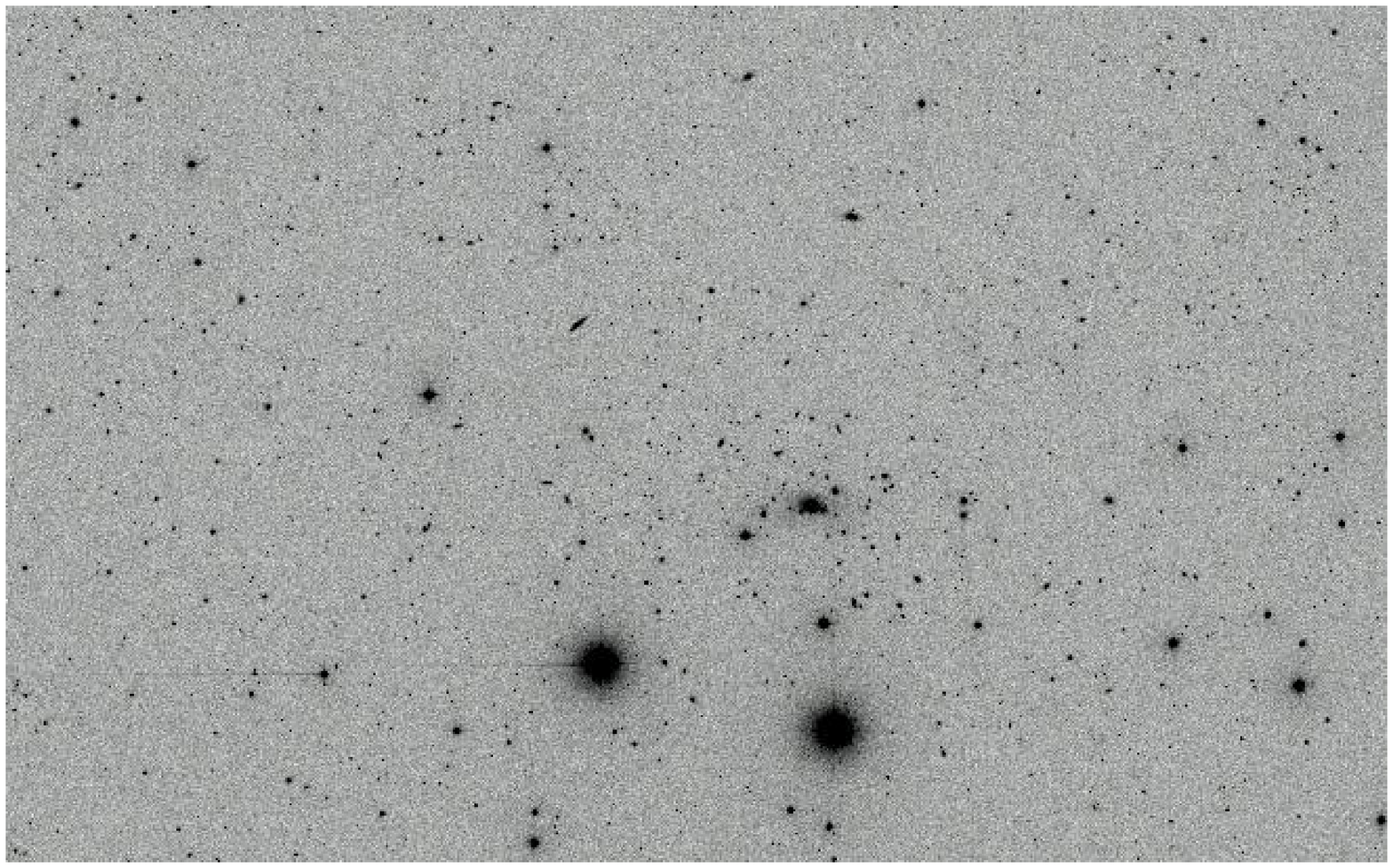}\hskip 0.1cm \plotone{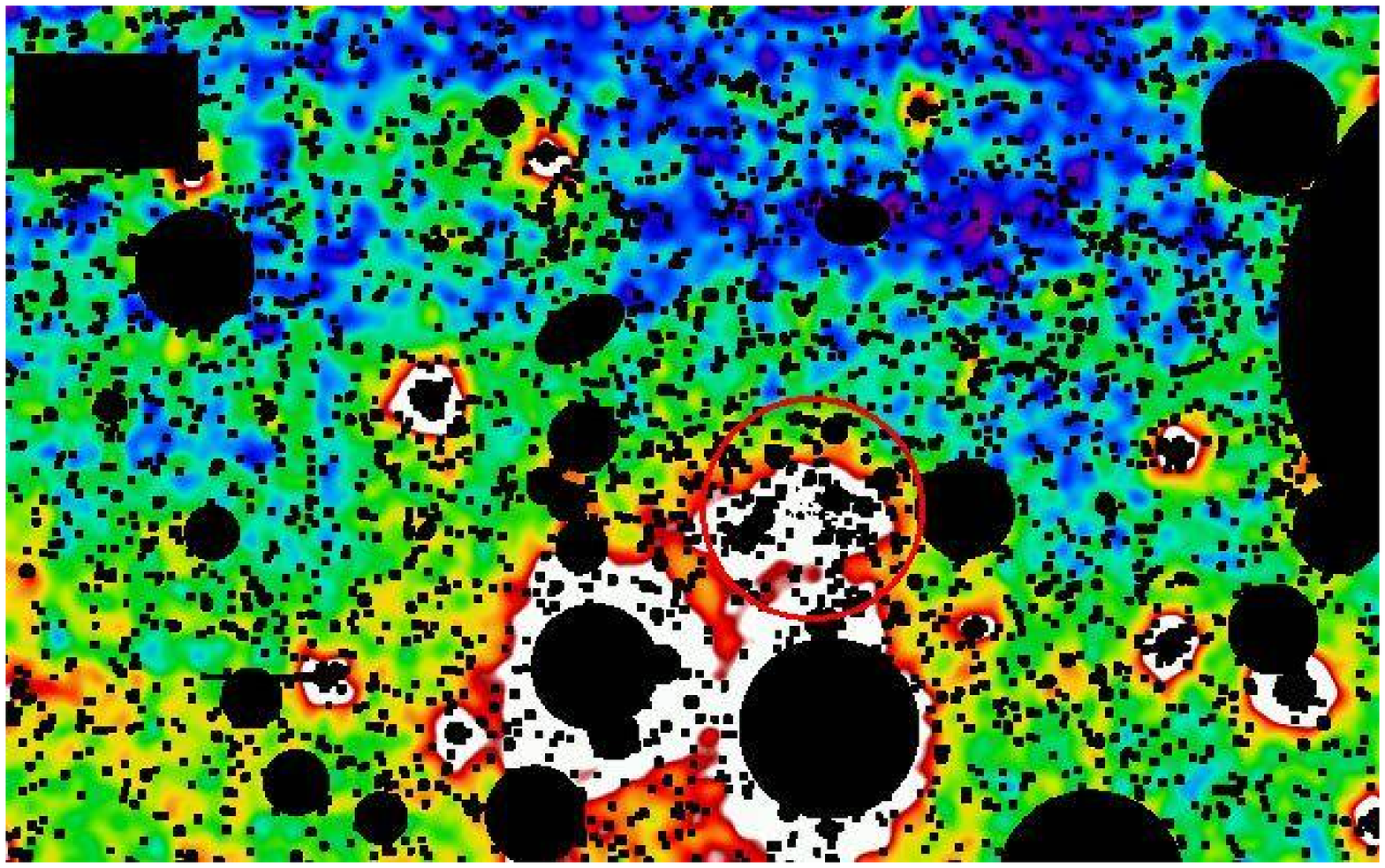}\\
\plotone{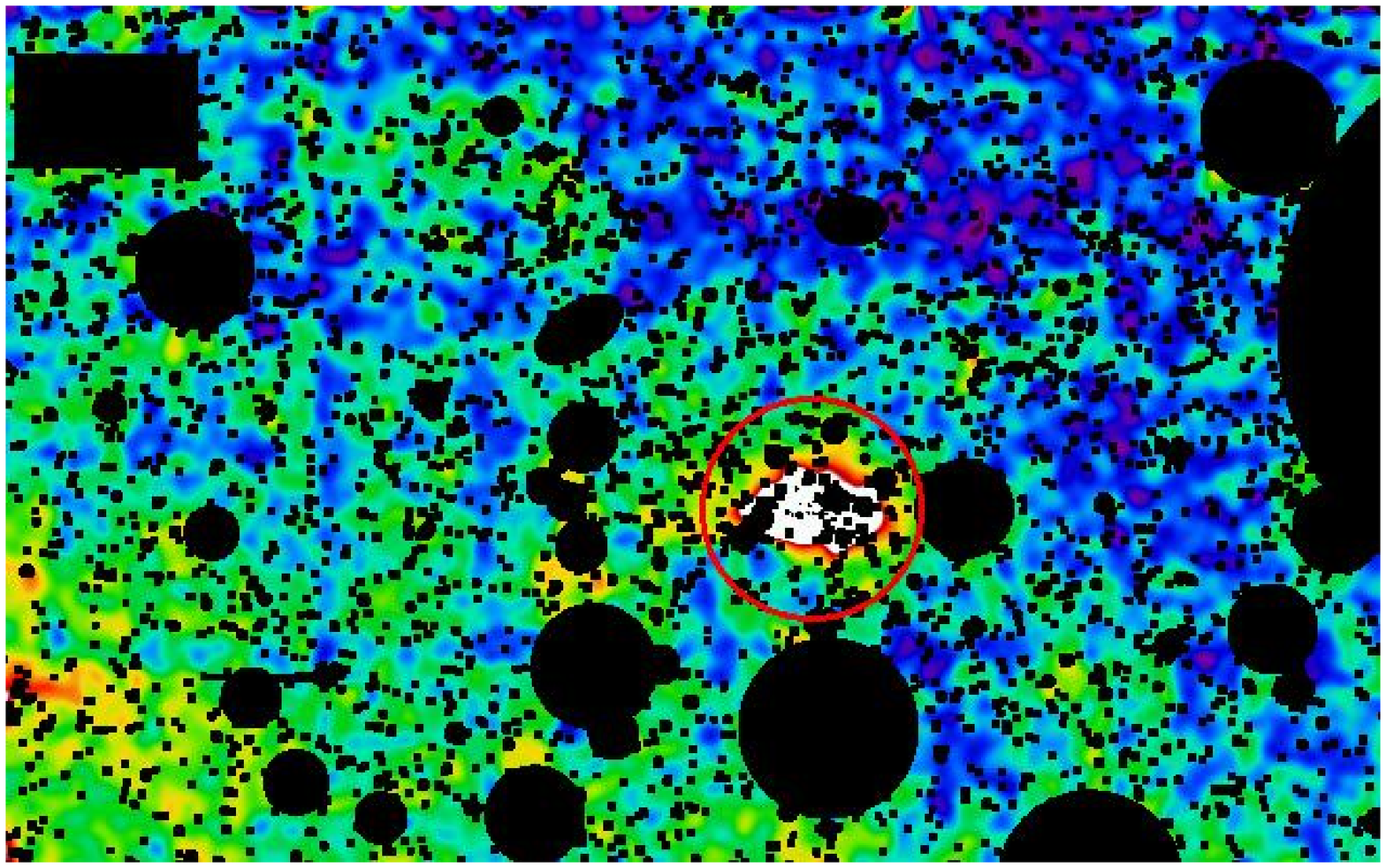}\hskip 0.1cm \plotone{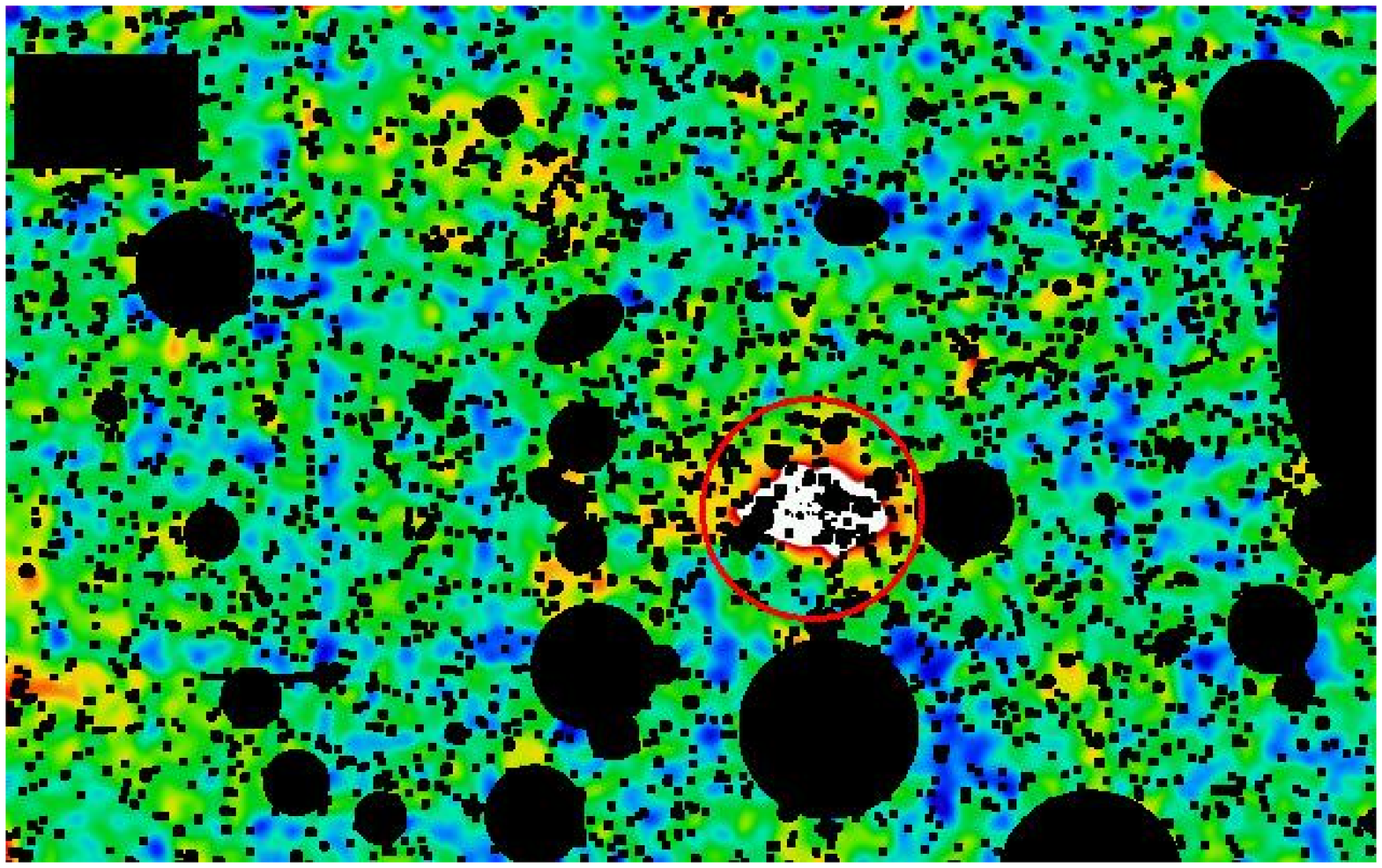}
\caption{Series of images of Abell 0122 illustrating the impact of PSF
subtraction and wavelet removal on the level of sky variation. The size of the
region shown is $30\arcmin\times19\arcmin$. The top left image is the data
immediately after co-adding, with the saturation level corresponding to
$\mu=21.8$ \sbu.  Subsequent images are Gaussian smoothed to enhance the
visibility of low surface brightness features and are each displayed with the
same contrast.  The saturation level in these images corresponds to $\mu=25.8$
\sbu.  We mask objects other than the BCG, which is marked by an overlaid
circle of radius $300$ \ho kpc (the size of the fitting region in \S
\ref{sec:profiles}).  The upper right image, like the upper left panel, is
constructed from the coadded individual scans. Significant excess flux is
visible along the perimeters of stellar masks. The lower left panel
demonstrates the improvement achieved with the PSF subtraction of saturated
stars. Finally, the lower right panel demonstrates the improvement achieved,
particularly in the uniformity of the background on large scales, by
subtracting the wavelet background map.}

\label{fig:reduction}
\end{figure*}

\subsubsection{Subtraction of Sky Variations: Second Pass}

After accounting for the PSFs of bright stars, we proceed with a second pass
to model the sky. The remaining large-scale gradients in the sky level may
arise from both instrumental (uneven illumination) and astrophysical (dust
emission) sources.  We construct a background map via wavelet decomposition of
the input image using the algorithm {\it wvdecomp} \citep{vikhlinin1998}.  We
use an \'a trous wavelet transform to detect all structures with $>2\sigma$
significance on scales of $2^0 - 2^9$ 
of $0.7\arcsec-358\arcsec$; see \citealt{starck1994} and references therein
for a detailed description of the \'a trous transform).  We subtract these
structures from the input image image and then generate the background map by
convolving the residual image with a kernel of scale $2^{10}$ pixels
($716\arcsec$). This scale is larger than the region within which the BCG
profiles are modelled (typically $r\la250\arcsec$). We find no evidence of
oversubtraction in the region of the BCG at the level of $\mu=27.3$ \sbu\ in
stacked background images.  The rms amplitude of the variations removed by
subtraction of the background maps is equivalent to $\mu\simeq26.8$ \sbu.

Figure \ref{fig:reduction} illustrates the impact of PSF subtraction and
wavelet removal of large scale gradients. The three masked images are shown at
the same contrast and are Gaussian smoothed to enhance low surface brightness
features.

\section{Modelling the Surface Brightness Profiles}
\label{sec:profiles}

We model the brightest cluster galaxy surface brightness profiles using
GALFIT, a two-dimensional galaxy-fitting program (Peng et al. 2002) that has
several advantages for our work. First, it permits simultaneous modelling of
multiple components. This feature enables us to model galaxies near the center
of the BCG that could otherwise bias the derived profiles, to model systems
with multiple dominant galaxies (in a subsequent study), and to model the BCGs
themselves with multiple components.  Second, because we fit the
two-dimensional luminosity distribution rather than the surface brightness
profile along the major axis, the $\chi^2$ of the fit is directly sensitive to
variations in the position angle and ellipticity with radius --- unlike many
previous studies.  The gradients provide important leverage in discerning
whether two component models for BCGs
\citep[e.g.,][]{schombert1986,schombert1988,porter1991} are superior to single
component \ser and \dev models.  Finally, the program generates an image of
the best-fit model, from which we calculate the model's radial profile using
the same masking as used for the data. This radial profile is valuable both
for visualization and comparison with previous studies.

There are five critical inputs to GALFIT, aside from the image itself, that
warrant detailed discussion. These inputs are the stellar PSF, the image mask,
the noise map, the mean sky level, and the parametric model for the surface
brightness distribution. Each is discussed below, and the sensitivity of the
results to the PSF, masking, and sky level are explored in the Appendix.

\subsection{PSF}
GALFIT requires a representation of the stellar PSF for its convolution with
the galaxy model. We derive the PSF using the IRAF implementation of DAOPHOT
II \citep{stetson1987}.  A Moffat model with first order residual variations
is generated for each image using the 150 brightest unsaturated stars.

\subsection{Masks}
Masks are generated using the procedure described in \S\ref{sec:data}. A key
difference is that here we do not automatically mask any objects within
$20\arcsec$ of the center of the BCG.  We visually inspect the images,
manually masking these objects when possible and leaving them unmasked for
modelling when necessary. We adopt the latter option only when part of the
object lies within $\sim 5\arcsec$ of the BCG.  Two clusters, Abell 2969 and
Abell 3705 require special attention.  In Abell 2969 two galaxies overlapping
the BCG were modelled, while in Abell 3705 a single large, bright galaxy
$2\arcmin$ away was modelled to minimize its impact upon the BCG profile at
large radii.

For each BCG, we also exclude all pixels outside a circular aperture of
physical radius 300 \ho kpc.  We choose a fixed physical radius to facilitate
comparison of the profiles, and choose 300 \ho kpc because this is roughly the
distance at which the surface brightness along the semi-major axis is
comparable to the $5\sigma$ sky level uncertainty for most BCGs in our sample.
For a typical BCG, $\sim25-45$\% of pixels within this physical aperture are
masked. In only three cases does the masking exceed 50\% of the pixels within
this radius: Abell 2376, Abell 2405, and Abell 3705, with $\sim55$\% of pixels
masked for each. For Abell 2376 and Abell 2405 the increased masking is due to
a nearby bright star; for Abell 3705 the cluster simply lies in a region of
high stellar density.

\subsection{Noise maps}
Noise maps are constructed directly from the data, which has zero mean sky due
to previous processing. Objects are initially masked and the
DIMSUM{\footnote{DIMSUM is the Deep Infrared Mosaicing Software package
developed by Peter Eisenhardt, Mark Dickinson, Adam Stanford, and John Ward.}
routine {\it iterstat} is used to directly compute the background rms noise
level.  Assuming Poisson statistics, we add the equivalent sky level to the
image and take the square root to generate a noise map, which is then smoothed
with a 2 pixel FWHM Gaussian. Smoothing provides a more robust measure of the
noise level, and use of a smoothing scale comparable to the seeing minimizes
spatial degradation of the noise map. While these noise maps yield model fits
that are good in absolute terms (reduced $\chi^2\simeq1$; \S
\ref{sec:results}), because these noise maps are approximate we only draw
conclusions from the relative $\chi^2$ values derived for the different
models.

\subsection{Sky Level}
\label{sec:profiles:skylevel}
The background level is a critical input because systematic error in the sky
will dominate the profile at large radii.  We use GALFIT to determine the sky
level, utilizing the same noise maps and masks as for the profile fitting.
The sky level is measured within circular annuli of radii $r=600-750$ \ho kpc
centered on the BCG. The extrapolated flux level of the central galaxy at
these radii is more than 3 mag fainter than it is at 300 \ho kpc (or
equivalently $<$7\% as bright), which is 2-3 times less than our $1\sigma$ sky
level uncertainty (see below), and hence negligibly impacts our estimate of
the sky level.

We use the ensemble properties of our sample to quantify the uncertainty in
the derived mean sky levels. The measured dispersion among the mean sky levels
observed for different galaxies ($\sigma_{obs}$) has two sources: true
differences in the global mean sky levels of different images ($\sigma_g$) and
our measurement uncertainty within the background annuli ($\sigma_m$).  If we
assume that these contributions add in quadrature, then
$\sigma_m^2=\sigma_{obs}^2-\sigma_g^2$.  To estimate $\sigma_g$ we measure the
global sky level over the entire image (excluding the BCG region) for each
cluster and then compute the dispersion.  We then measure the mean sky level
within the background annulus for each BCG and compute $\sigma_{obs}$. In both
cases we use GALFIT to determine the mean sky level, and together these data
enable us to derive $\sigma_m$, the $1\sigma$ uncertainty in the sky level.
For a typical galaxy in our sample this $1\sigma$ uncertainty corresponds to
$\mu\simeq28.4$ \sbu.  {\footnote{The precise value varies depending on the
photometric zeropoint of the field, because the calculation is performed using
uncalibrated counts.}}  In one case, Abell 2376, we increase our estimate of
the sky uncertainty by a factor of two to account for the impact of a
proximate bright star.

\subsection{Input Models}
\label{sec:results:models}
We test three different functional forms for the BCG surface brightness
distribution.  Two of these are single component 
\epsscale{1.0}
\begin{figure*}
\plotone{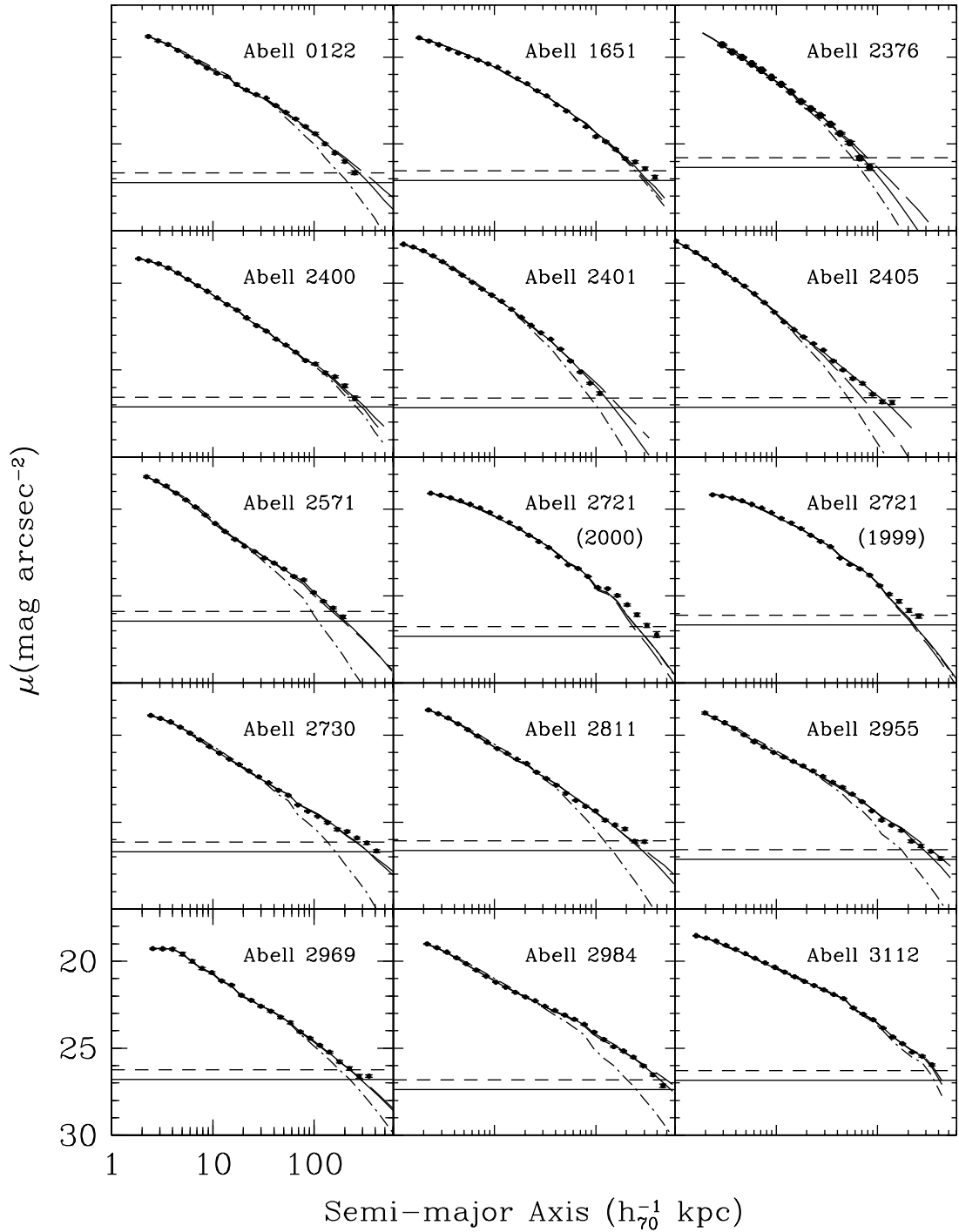}
\caption{$I-$band surface brightness profiles for the 24 BCGs modelled in
this paper. The error bars, which are typically smaller than the symbols, are
statistical. Overlaid are the corresponding surface brightness profiles for 
the best-fit two-dimensional models. We emphasize that these are not fits to
the one-dimensional profile, but rather measurements of the two-dimensional
models taken in the same apertures as for the data. The
\dev profile (short-long dashed) is a poor representation of the data in most
cases, while the $r^{1/n}$ (dotted) and \ddev (solid) profiles provide
improved fits. The solid and dashed horizontal lines correspond to the
$3\sigma$ and $5\sigma$ sky level uncertainties, respectively.}
\label{fig:profiles}
\end{figure*}
\addtocounter{figure}{-1}
\begin{figure*}
\plotone{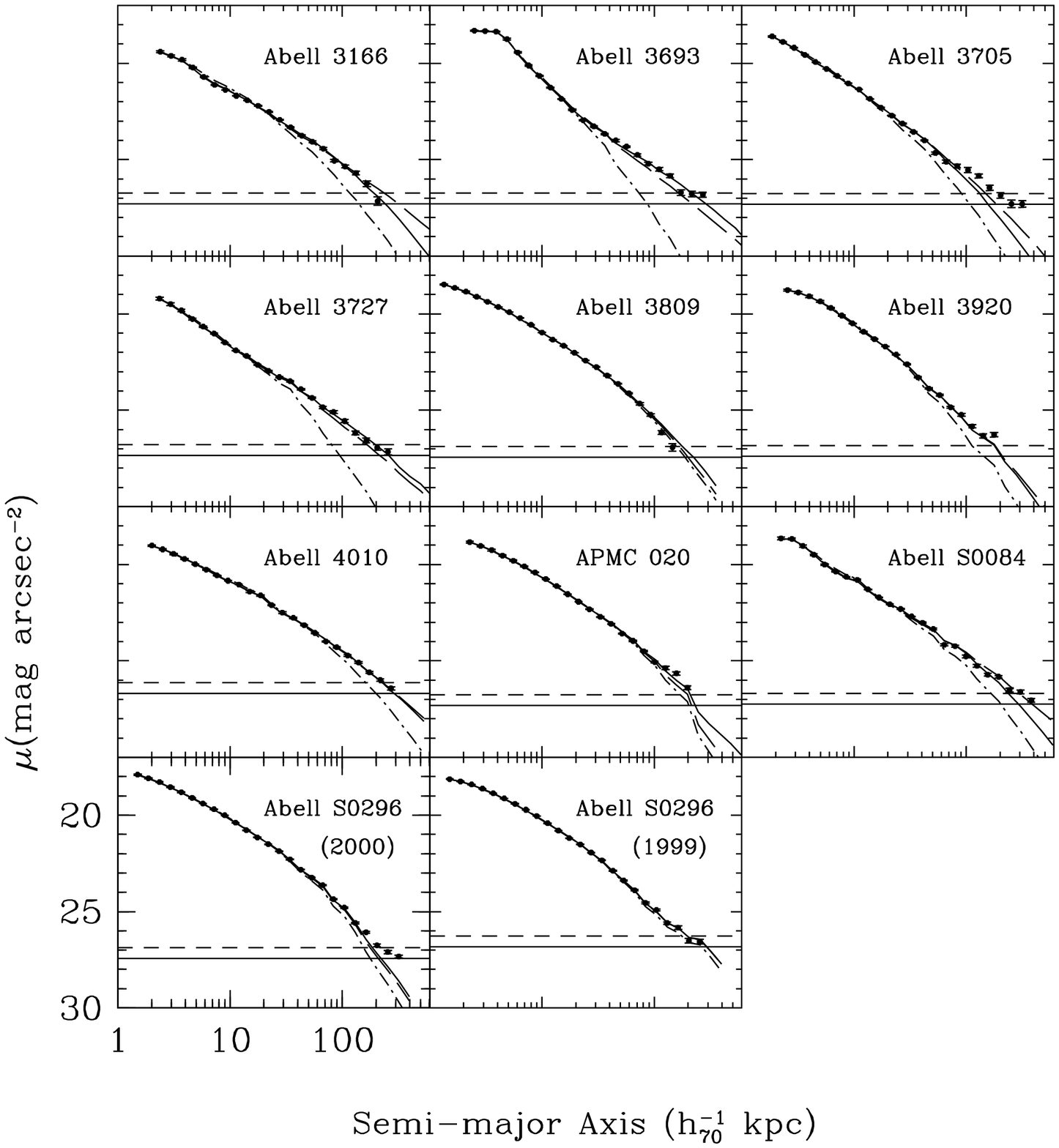}
\caption{Continued.}
\end{figure*}

\epsscale{1.0}
\begin{figure*}
\plotone{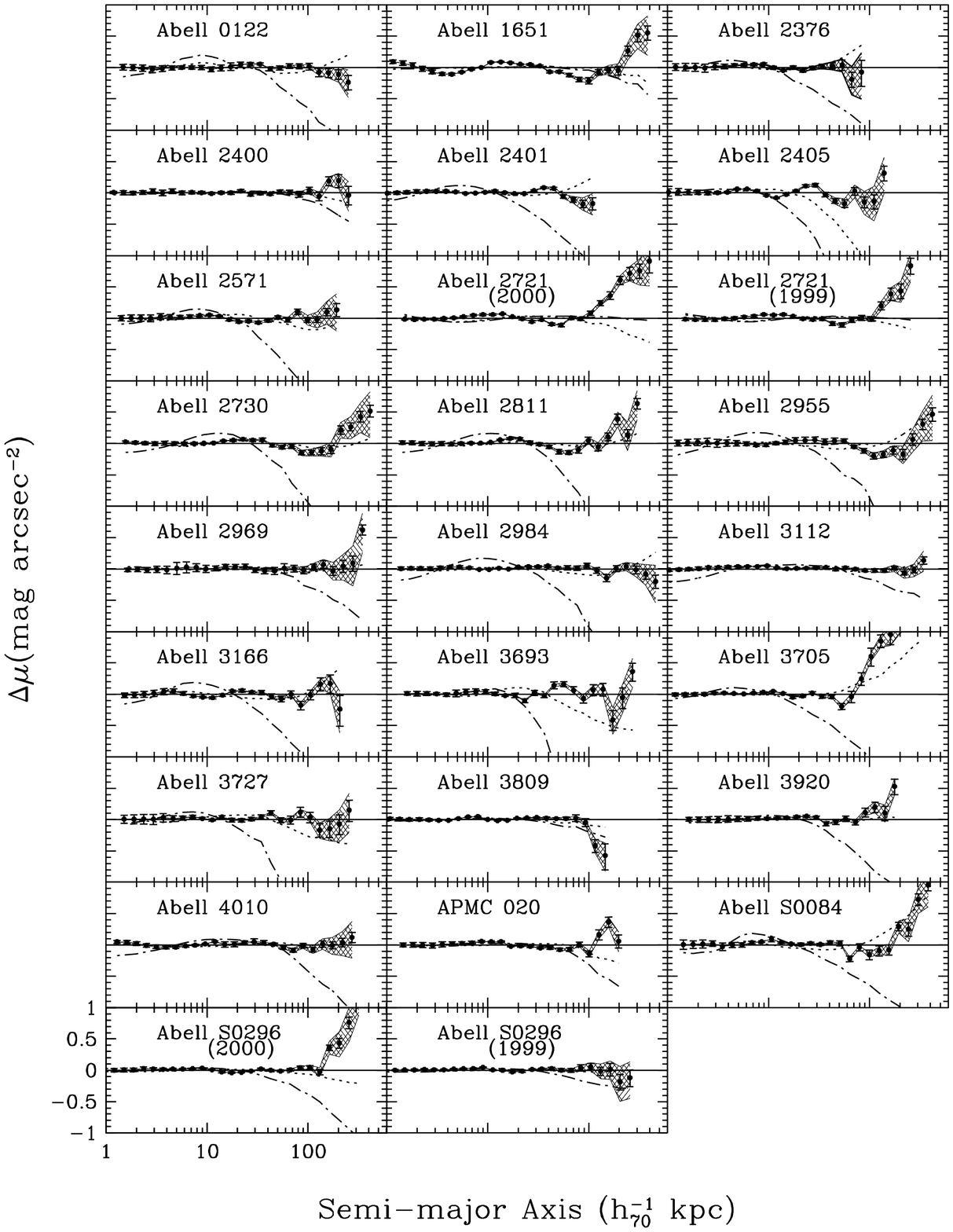}
\caption{Surface brightness residuals relative to the \ddev model. The
line types are the same as in Figure \ref{fig:profiles}, but we have now also
added a shaded region denoting the $1\sigma$ systematic sky level
uncertainty. We note that in two cases, Abell 2721 and Abell 3705, bright saturated
stars appear to skew the residuals at large radii. In both cases the stellar PSFs
were subtracted and the central regions masked, but the systematic sky level uncertainty
is likely larger than our estimate in these cases.}
\label{fig:residuals}
\end{figure*}

\epsscale{1.0}
\begin{figure*}
\plotone{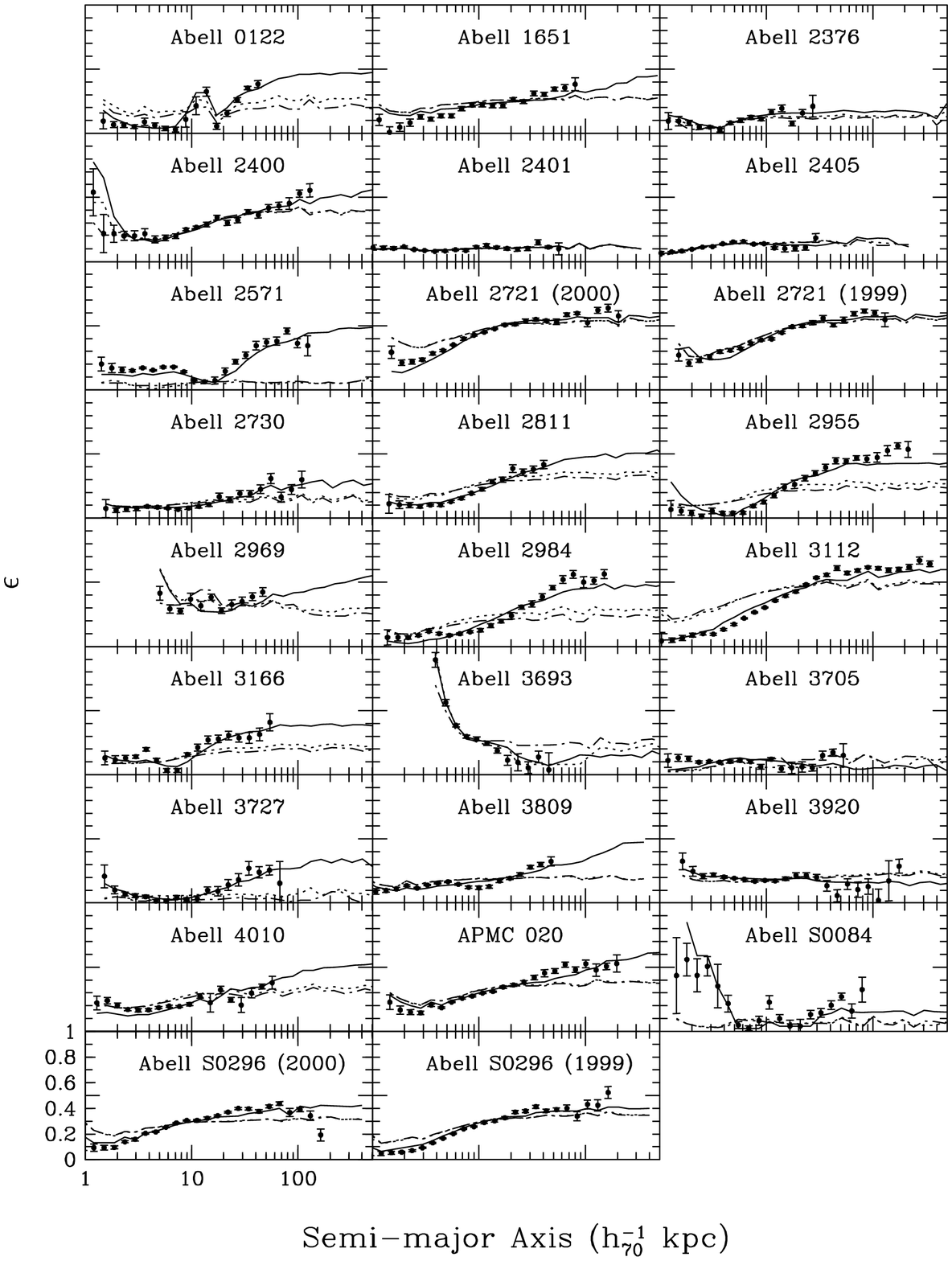}
\caption{Ellipticity profiles for the 24 BCGs modelled in
this paper. The line types are the same as in Figure \ref{fig:profiles}. Note
that the sharp, non-monotonic changes in ellipticity and position angle seen
in a few of the profiles (e.g.  Abell 0122) occur when a significant fraction
of the data at that radius are masked. These features are thus more a function
of the masking than indicative of the profiles, but are produced in both the
model and data profiles since the same masking is applied in both cases. In
contrast, large monotonic changes in the ellipticity (e.g. Abell 3112) are
physical in origin.  The \ddev profile is best able to reproduce these large
monotonic changes.}
\label{fig:ellipticity}
\end{figure*}

\epsscale{1.0}
\begin{figure*}
\plotone{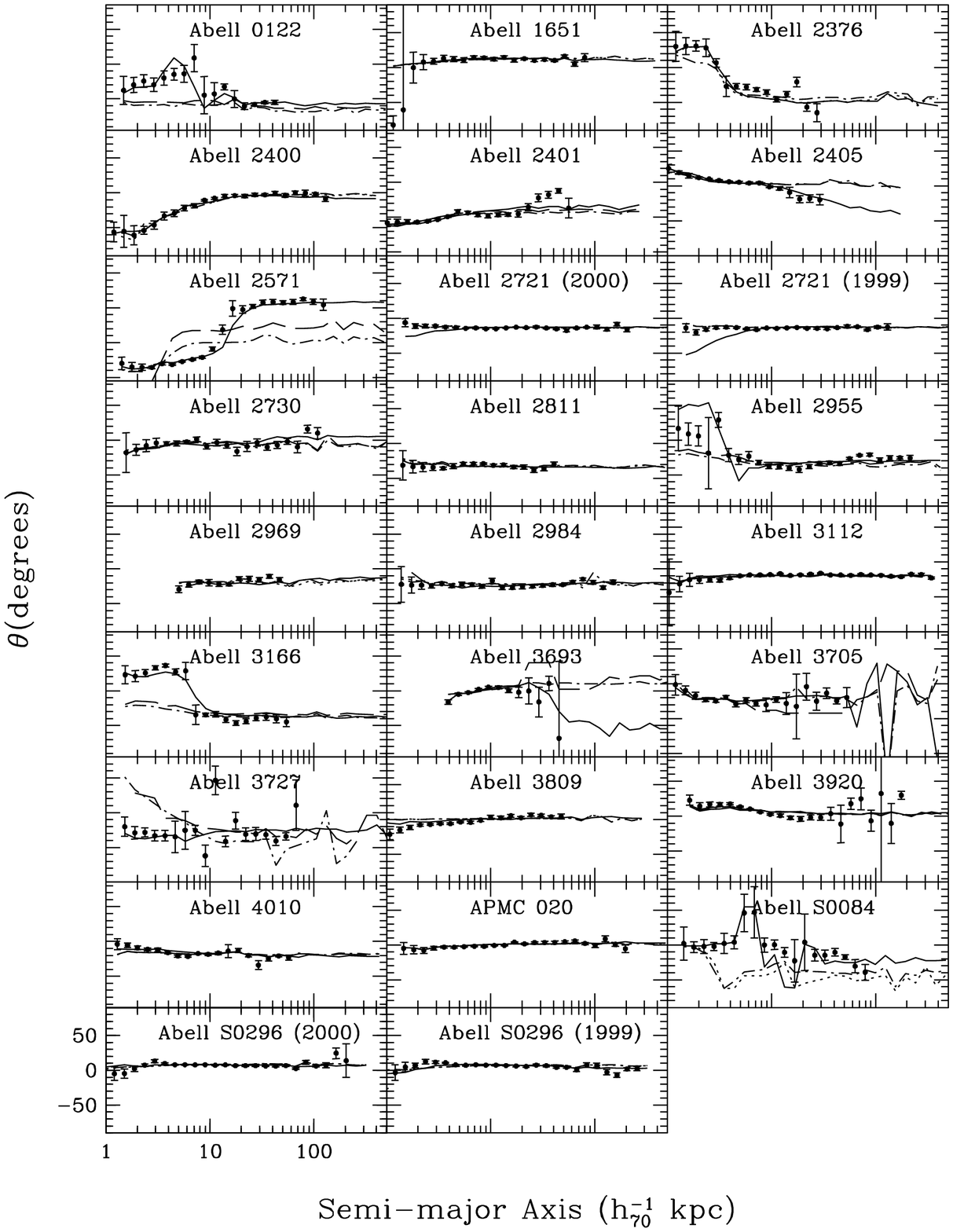}
\caption{Position angle as a function of radius. The line types are the same as in 
Figure \ref{fig:profiles}. The zeropoint for $\theta$ is arbitrary for each galaxy,
and is defined to avoid having $\theta$ wrap around the vertical axis.
Only the \ddev profile can reproduce the large,
rapid position angle changes seen in some of the profiles.}
\label{fig:pa}
\end{figure*}

\noindent deVaucouleurs ($r^{1/4}$) and
\ser ($r^{1/n}$) models, while the third, in the spirit of
\citet{schombert1988}, is a two component model consisting of a pair of \dev
profiles with independent parameters.  We will refer to the latter model as
the \ddev model henceforth.  The $r^{1/n}$ model has one more degree of
freedom (dof) than the \dev model, while the second component in the \ddev
model adds six dof relative to the single \dev model
($x$,$y$,$\mu_e$,$r_e$,$\theta$,$\epsilon$).  For each functional model we run
GALFIT with a series of different initial parameters to ensure convergence to
the global minimum for each model, and iterate upon the solution until $\Delta
\chi^2<10^{-3}$.

\vskip 1.0cm
\section{Results and Discussion}
\label{sec:results}

\subsection{Evidence for Two Components}
\label{subsec:twocomponents}
The optimal parameters for each model and the $\Delta\chi^2$ values between
different models are presented in Tables
\ref{tab:devparams}---\ref{tab:ddparams}.  In these Tables we include both the
statistical parameter uncertainties from GALFIT and, in parentheses, the
uncertainties corresponding to 1$\sigma$ changes in the sky level (see \S
\ref{sec:profiles:skylevel} and the Appendix).  To illustrate the quality of
the models, we also present one-dimensional profiles extracted using the IRAF
task {\it ellipse}.  In Figure \ref{fig:profiles} we present surface
brightness profiles along the semi-major axis, and in Figure
\ref{fig:residuals} we show the residuals relative to the \ddev model. In both
figures we employ the same masking and elliptical apertures for the data and
models to enable direct comparison.{\footnote{A potential concern is that the
lack of ellipticity and PA gradients for the \ser and \dev models could bias
the residuals shown in Figure \ref{fig:residuals}. We have performed tests to
assess the importance of this bias, and find that the effect is negligible.}
The model images output by GALFIT include convolution with the input PSF, and
hence comparison to the data is valid over the entire radial range shown.  In
Figure \ref{fig:residuals} we overlay shaded regions corresponding to the
1$\sigma$ systematic sky level uncertainty
(\S\ref{sec:profiles},\S\ref{sec:robustness:sky}) to clarify the significance
of deviations between the data and models.

Figures \ref{fig:ellipticity} and \ref{fig:pa} show the ellipticities and
position angles derived for both the data and best-fit models. While the
single component models have fixed values for both quantities, there are
gradients visible in the Figures for these models because we have folded into
the profiles several additional factors to enable direct comparison with the
data.  First, masking of galaxies near the center of the BCG can lead to
variations in both quantities. We use the same masking for both the data and
models to ensure a fair comparison.  Second, seeing acts to decrease the
central ellipticity and can cause position angle changes if the ellipticity is
small. The models are convolved with the PSF from the image and so exhibit the
same decrease as the data. Third, in some instances we found it necessary to
model rather than mask secondary galaxies near the center of the BCG, which
are listed in the caption for Figure \ref{fig:ellipticity}. The flux from
these galaxies is included in the plotted profiles for both the data and
models.  Abell 2400 is a good example of the impact that such a companion can
have on the central profile.  Together the above factors explain the gradients
seen in the single component models; yet, even including these factors, these
models cannot reproduce the ellipticity and position angle variations observed
in some of the BCGs.

\subsubsection{\ser versus \dev}

Our data demonstrate that the $r^{1/n}$ profile provides a dramatic
improvement relative to the \dev profile.  The additional degree of freedom
decreases $\chi^2$ by an average of 3649 (Table \ref{tab:serparams}), and
substantially improves the fit at large radii (Figures
\ref{fig:profiles}---\ref{fig:residuals}). In only two cases is the decrease
in $\Delta\chi^2$ minimal (Abell 1651 with $\Delta\chi^2=38$ and Abell 2721
with $\Delta\chi^2=19-84$ for data from two different years).  Abell 1651 is
the cluster from our original pilot study in which we found that the BCG
showed no strong evidence of departure from \dev \citep{gonzalez2000}. We list
in Table \ref{tab:a1651} the \ser parameters derived in both studies of this
BCG, noting that the results are consistent despite different modelling
procedures and bandpasses.

\subsubsection{\ddev versus \ser}
Although the \ser profile is superior to the \dev model for our BCG sample and
provides a fit that is good in absolute terms ($\chi^2_\nu\simeq1$), the
$\Delta\chi^2$ values in Table \ref{tab:ddparams} argue that the \ddev profile
provides the best parametric fit to the data. With five additional degrees of
freedom, the \ddev profile decreases $\chi^2$ by an average 1039 relative to
the \ser profile, which is highly significant.  The reasons that the \ddev
model provides a superior fit can be assessed by inspection of Figures
\ref{fig:profiles}---\ref{fig:pa}. For roughly a third of the galaxies the two
components are required to reproduce the observed large gradients in the
ellipticity and/or position angle. In the other two-thirds, where these
gradients are small, the fit is still statistically superior. The \ddev model
is thus preferred due to both its ability to fit the mean surface brightness
profile and improved ability to reproduce the gradients.

\subsubsection{Physical interpretation of \ddev model}
Does the success of the \ddev model imply that there are two physically
distinct components?  Perhaps the ellipticity gradients and isophotal twists
are simply the result of viewing triaxial galaxies off-axis. If so, then the
individual components are not physical but rather a convenient means of
approximating a triaxial stellar distribution.  There are several reasons to
conclude that this is not the case.  First, there are radial regions over
which the position angle and ellipticity are constant (e.g. Abell 2571), which
implies that at those radii we are either viewing the system along a principal
axis or else that the system is not triaxial (see Figures
\ref{fig:ellipticity}---\ref{fig:pa}). If the former, then the ellipticity and
position angle should remain constant at all radii.  Second, the more dramatic
position angle shifts (e.g. Abell 2571, Abell 3166) happen very suddenly,
which again should not be the case for a triaxial system.  Third, the apparent
position angle of a triaxial body is in general different than the projected
position angle of the true major axis.  The known correlation between BCG and
cluster orientation \citep{binggeli1982} argues against these galaxies being
strongly triaxial systems viewed at random angles.  Fourth, if we are viewing
projected triaxial ellipsoids, then we should expect all the galaxies to show
position angle swings, as is true with most ellipticals. Many of the BCGs in
this sample in fact lack significant position angle gradients, and it is
statistically quite improbable that we are viewing such a large fraction along
their principal axes.  Finally, for a triaxial system the relative importance
of the apparent second component should be a strong function of viewing angle
(i.e. when the BCG is observed near the principal axis this component should
be negligible).  We find instead that the scale lengths of the two components
are strongly correlated (see \S\ref{sec:correlations}).

Aside from the above arguments against the observed gradients in ellipticity
and position angle being due to triaxiality, ancillary data also favor the two
components having physical significance. Most compelling are the velocity
dispersions in two nearby clusters, Abell 2029 \citep{dressler1979} and Abell
2199 \citep{carter1999,kelson2002}, which are observed to rise towards the
cluster velocity dispersion at large radii.  Such a rise is strong evidence
that the orbits of the outer stars are associated with the gravitational
potential of the cluster rather than the central galaxy.  Taken with the
arguments against triaxiality, these data argue that the success of the \ddev
model is not simply a result of analyzing triaxial bodies, but rather that
these two components are indeed distinct physical entities.

Finally, we emphasize three additional points. First, while we model the two
components as a pair of \dev profiles, we are {\it not} specifically
concluding that the two components are \dev in nature. This model simply
provides an adequate description of our data with a minimum number of free
parameters. The outer component could conceivably be better described by an
NFW profile \citep{navarro1996,navarro1997}, but we have not yet explored this
model.  Second, for any single BCG, we caution the reader to not place undue
significance on the structural parameters of the individual components in the
\ddev model. While the results are robust for the statistical sample, the
decomposition for a given BCG may be biased by internal surface brightness
structures or tidal features.  Third, we are not claiming to see the dramatic
profile breaks characteristic of classical cD envelopes
\citep[e.g.][]{schombert1986,schombert1988} as ubiquitous features. The
transition between the two components in our data is typically much more
subtle.

\epsscale{1.0}
\begin{figure}
\plotone{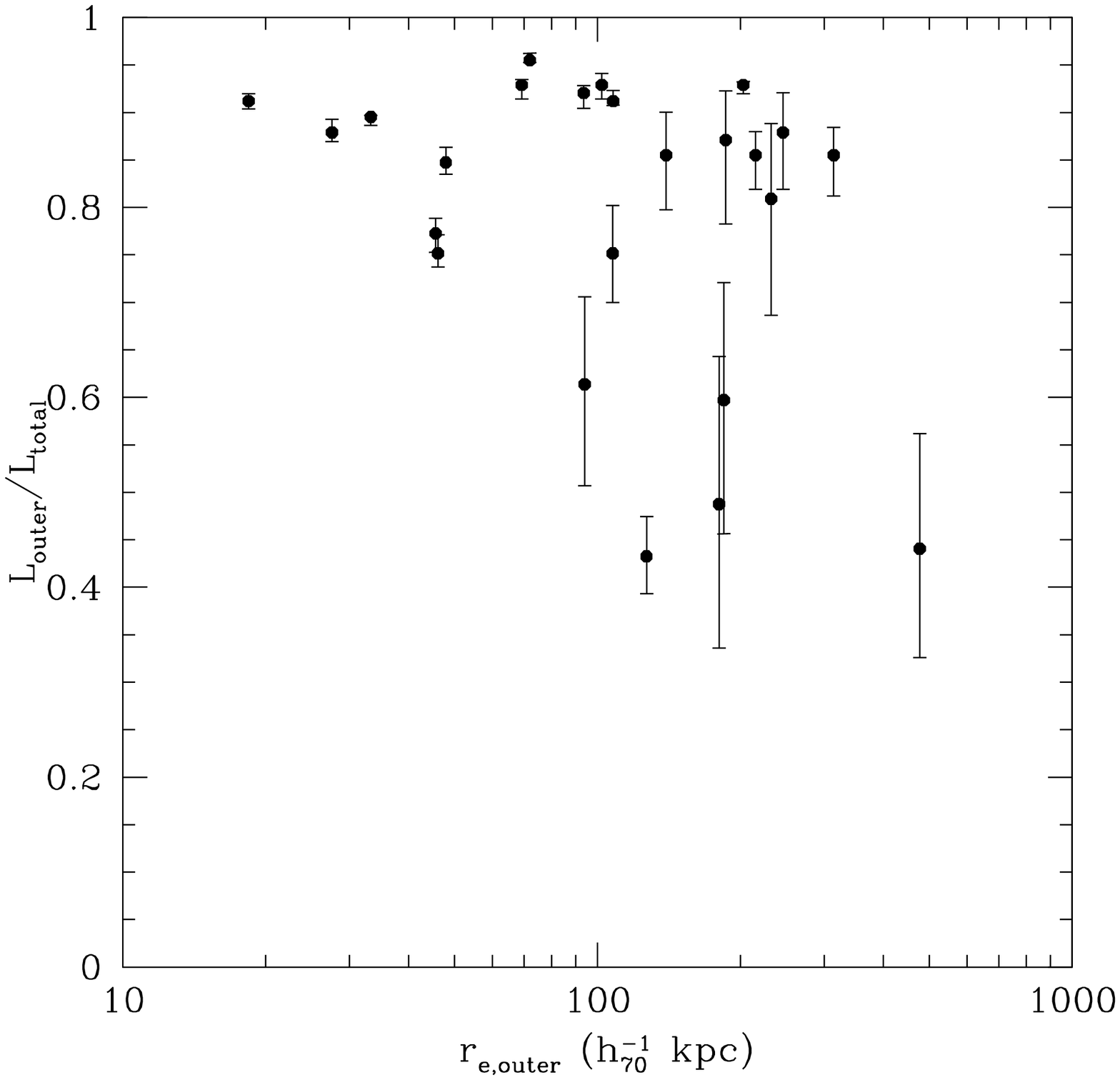}
\caption{Fractional luminosity contribution of the outer component, where
$L_{total}$ is defined as the combined luminosity of the inner and outer 
components. The outer component in most cases contains $>$80\% of the 
luminosity.}
\label{fig:mrat}
\end{figure}

\begin{figure*}
\plottwo{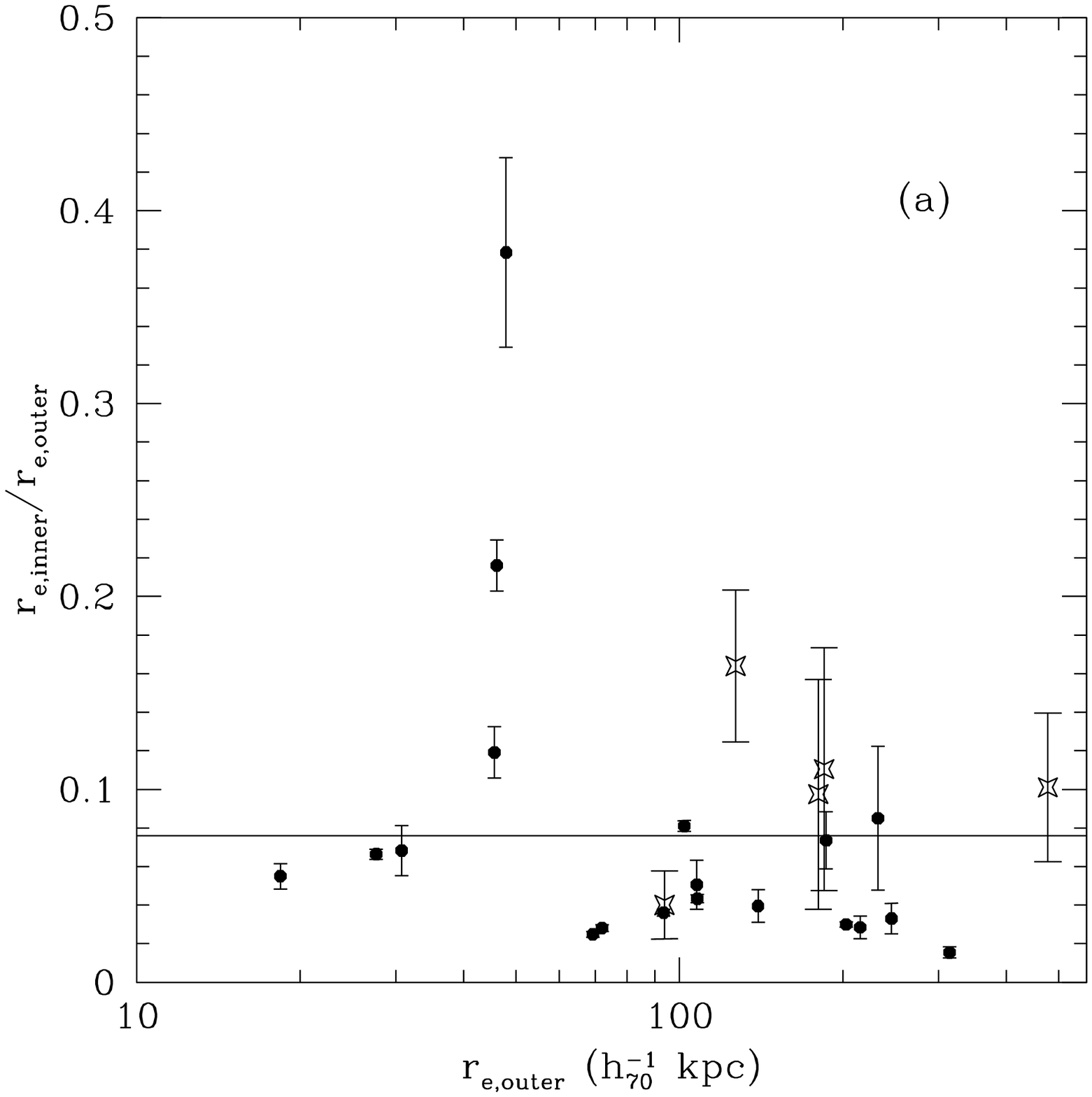}{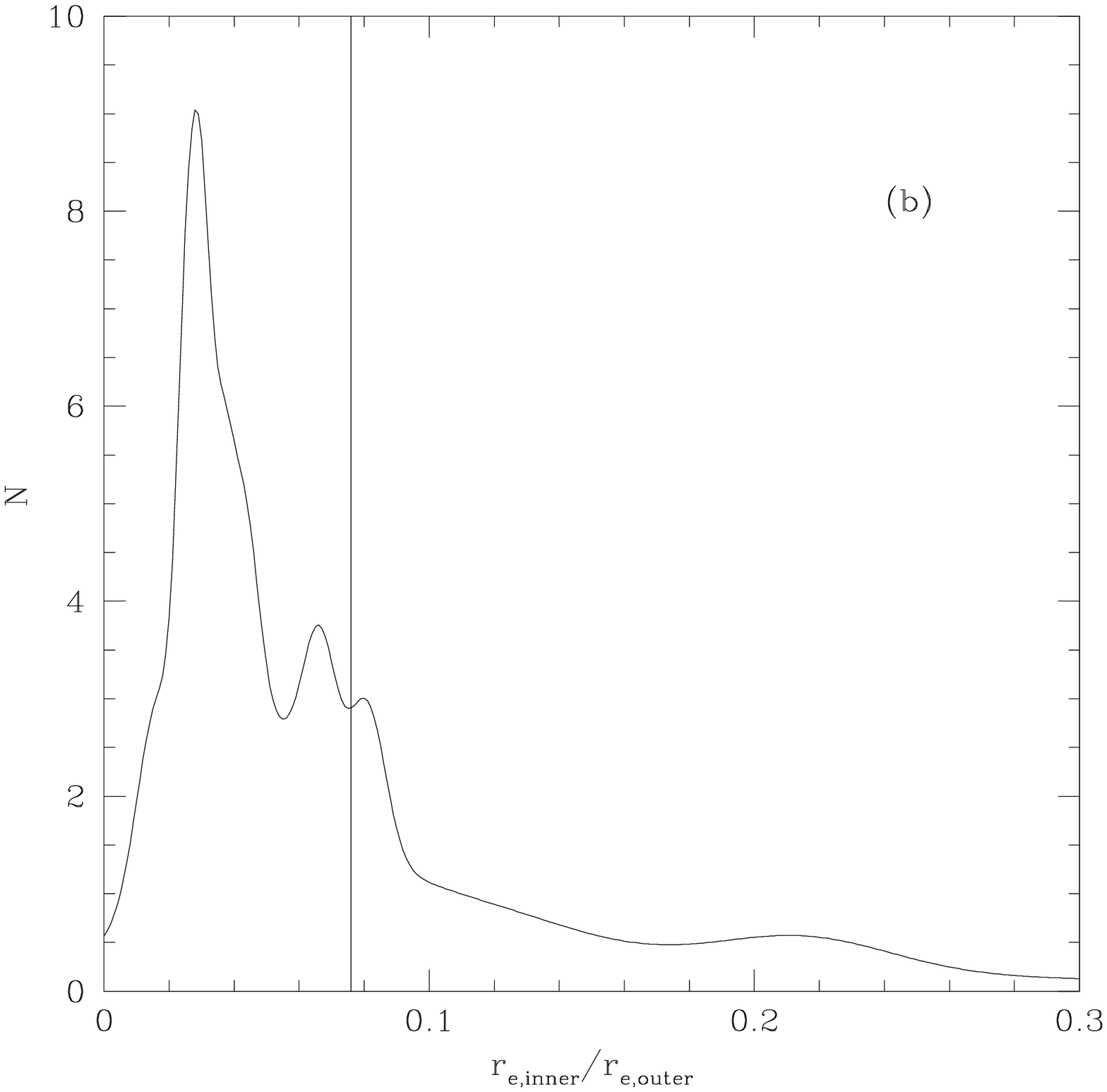}
\caption{$(a)$ Ratio of the effective radii for the inner and outer
components. The open starred points correspond to galaxies in which the outer
component contains less than two thirds of the total luminosity (see Figure
\ref{fig:mrat}), and the solid line corresponds to the ratio between the
effective radii of the stellar and dark matter distributions in the BCG
simulation of \citet{dubinski1998}.  The highest point is the BCG in Abell
2721. For this galaxy both the position angle and axis ratio differences
between the two components are small ($|\Delta\theta|<10\degr$ and $\Delta
b/a<0.1$), limiting our leverage in separating the two components.  $(b)$
Smoothed distribution of $r_{e,inner}/r_{e,outer}$. Each galaxy is represented
as a gaussian with $\sigma=2
\times(\sigma_{statistical}^2+\sigma_{sky}^2)^{1/2}$. The factor of 2 is an
arbitrary scaling factor introduced to yield a smooth distribution. The solid
line is the same as in $(a)$.  The distribution is strongly
peaked at $r_{e,inner}/r_{e,outer}\simeq35$, suggesting that the two physical
scales are correlated.  }
\label{fig:r12}
\end{figure*}

\begin{figure*}
\plottwo{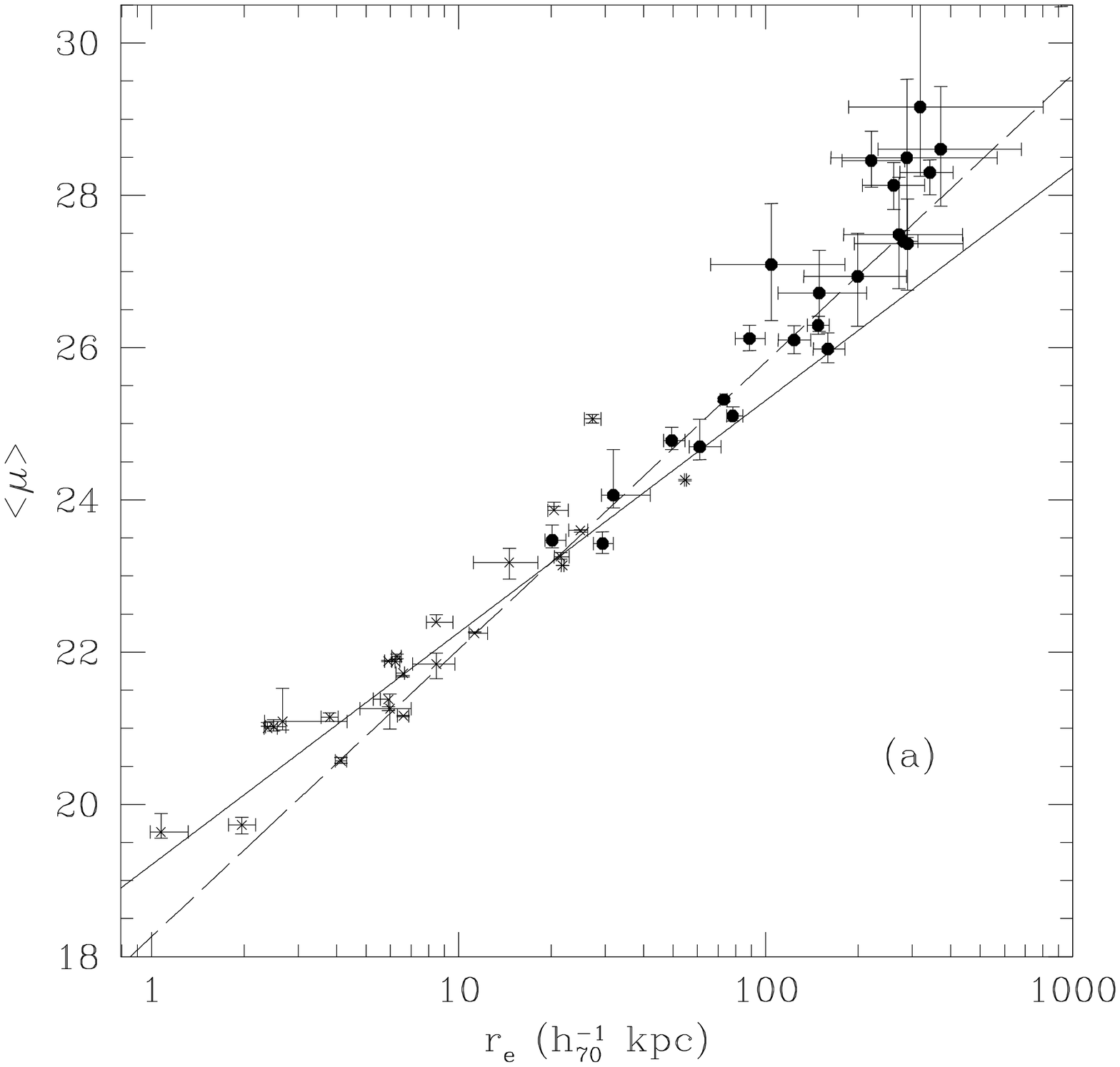}{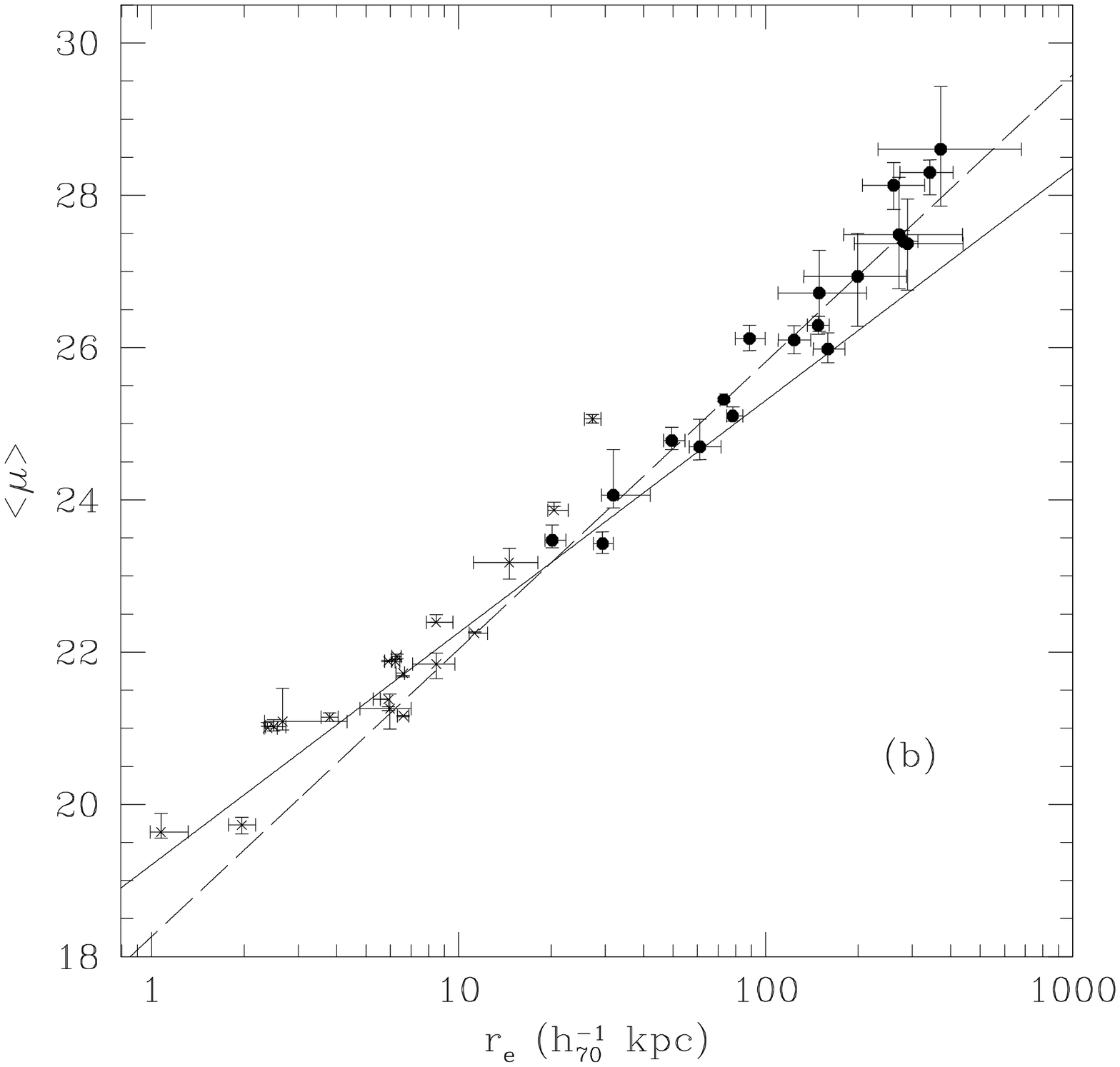}
\caption{Surface brightness vs effective radius for the \ddev model. Crosses
and filled circles denote the inner and outer components, respectively. The
lefthand panel $(a)$ shows all galaxies, while the righthand panel $(b)$ shows
only those with $>$2/3 of the total luminosity in the outer component.  The
solid line ($<I> \propto r_e^{-1.22}$) corresponds to the relation for
ellipticals from \citet{jorgensen1996}.  The dashed line ($<I> \propto
r_e^{-1.5}$) is a best fit to the outer component data in $(b)$.
We note that the inner component data point that lies well above the
canonical relation corresponds to Abell 2721, which also has an anomalously
large ratio of inner to outer effective radii (see Figure \ref{fig:r12}). For
Abell 2721 both the position angle and axis ratio differences between the two
components are small ($|\Delta\theta|<10\degr$ and $\Delta b/a<0.1$), limiting
our leverage in separating the two components. We conclude that the inner and
outer components both obey tight $<\mu>-r_e$ relations. The slope of the
relation for the inner component is comparable to the projection of the
elliptical fundamental plane, while the slope for the outer component is
steeper.  Note that in this and all subsequent figures that plot parameter
correlations the true parameter uncertainties are correlated and should
formally be plotted as correlated error ellipsoids. The error bars that we
plot correspond to the $1\sigma$ uncertainties in each parameter, derived by
adding in quadrature the statistical and systematic sky level uncertainties.
}
\label{fig:mueredd}
\end{figure*}

\subsubsection{Evidence for additional components}
Given that we find a two component model to be statistically superior, one can
ask whether there is evidence for any additional components. The residuals in
Figure \ref{fig:residuals} indicate that in a few cases the flux exceeds the
best-fit model by several sigma in the outermost few radial bins. In two of
these cases, Abell 2721 and Abell 3705, the likely origin of this excess is
residual flux from saturated stars at the corresponding radii.  In another,
Abell S0296, only the 2000 observations show an excess, indicating that the
origin is likely instrumental rather than physical. The other two cases with
$\sim2\sigma$ excesses (Abell 1651 and Abell S0084) lack a similar
explanation, and we also note that the sample as a whole exhibits a slight
tendency for excess flux at large radii. With the current data, however, we
cannot discern whether these excesses are due to a systematic error or a true
physical excess relative to the model, arising from either an additional
component or the outer profile having a form other than \dev.

\subsection{Relative Properties of the Two Components}

\subsubsection{Luminosities}
\label{subsec:luminosities}
What fraction of the light does the outer component contribute?  Figure
\ref{fig:mrat} demonstrates that the outer component typically contains
$\sim80-90\%$ of the total light from both components. In only five cases is
the outer component's luminosity $<$70\% of the total, and those systems tend
to be the ones with the largest uncertainties in the magnitude of the outer
component (e.g. Abell 1651, Abell 2400, Abell 2405, APMC 020).  In subsequent
plots we mark these five systems with starred symbols to illustrate the impact
of this systematic difference in luminosity ratio (whether real or due to an
error in decoupling the two components).  

\subsubsection{Effective Radii}
How do the effective radii of the two components compare and are they
correlated?  The range of effective radii for the inner component is $1-48$
\ho kpc, while the range for the outer component is $18-480$ \ho kpc.  Figure
\ref{fig:r12} shows that the effective radii of the two components are
coupled, with a typical ratio $r_{e,outer}/r_{e,inner}\sim 10-40$. For
comparison, we overlay a line denoting the ratio of the projected half-mass
radii for the dark matter and stellar components ($r_{DM}/r_{stellar}=13.2$)
from the cluster simulation of \citet{dubinski1998}. This simulation recovered
a purely \dev BCG profile, with a scale comparable to that of the inner
components observed in this study, but the stellar component was not modelled
with sufficient resolution to produce excess light at large radii.

\subsubsection{Fundamental Planes}
Do the individual components in the \ddev model independently obey a
$<\mu>-r_e$ relation similar to the projection of the elliptical galaxy
fundamental plane?  We find that both components do indeed form tight
$<\mu>-r_e$ relations, as seen in Figure \ref{fig:mueredd}$a$.  The slope of
the relation for the inner component is consistent with the elliptical
fundamental plane, which has $<I> \propto r_e^{-1.22}$
\citep{jorgensen1996}. In constrast, the slope of the relation for the outer
component is steeper than for elliptical galaxies. If we exclude systems in
which the outer component contributes less than two-thirds of the total
luminosity (see \S \ref{subsec:luminosities} above), the slope of the relation
is $<I> \propto r_e^{-1.5}$ and the scatter is consistent with the
observational uncertainties (Figure \ref{fig:mueredd}$b$).

\subsubsection{Ellipticities and Position Angles}
Two other parameters of basic interest are the ellipticities and relative
offsets in position angle of the two components.  The outer component is
generally more elliptical than the inner component (Figure
\ref{fig:paellcorr}$a$), consistent with previous observations that found
increasing ellipticity with radius \citep[c.f.][]{porter1991}.  The
ellipticity of the outer component is comparable to typical observed
ellipticities for the distribution of cluster galaxies
\citep{carter1980,plionis1991,buote1996,basilakos2000}.

The inner and outer components are closely aligned ($\Delta\theta < 10\degr$)
for roughly 40\% of our sample, indicating that both may retain information
about the initial infall direction of the system (Figure
\ref{fig:paellcorr}$b$ ).  Those inner components that are not well-aligned
show a preference for large $\Delta\theta$ ($\Delta\theta > 45\degr$ with a
distribution that rises toward $90\degr$). This offset may be the result of
infall events along other directions that determine the current position angle
of the central galaxy.  It is also possible that this could be a simple
projection effect in some cases --- particularly cases in which $\Delta\theta
\sim 90\degr$ and $b/a>0.9$ (e.g. Abell 3727).

\begin{figure*}
\plottwo{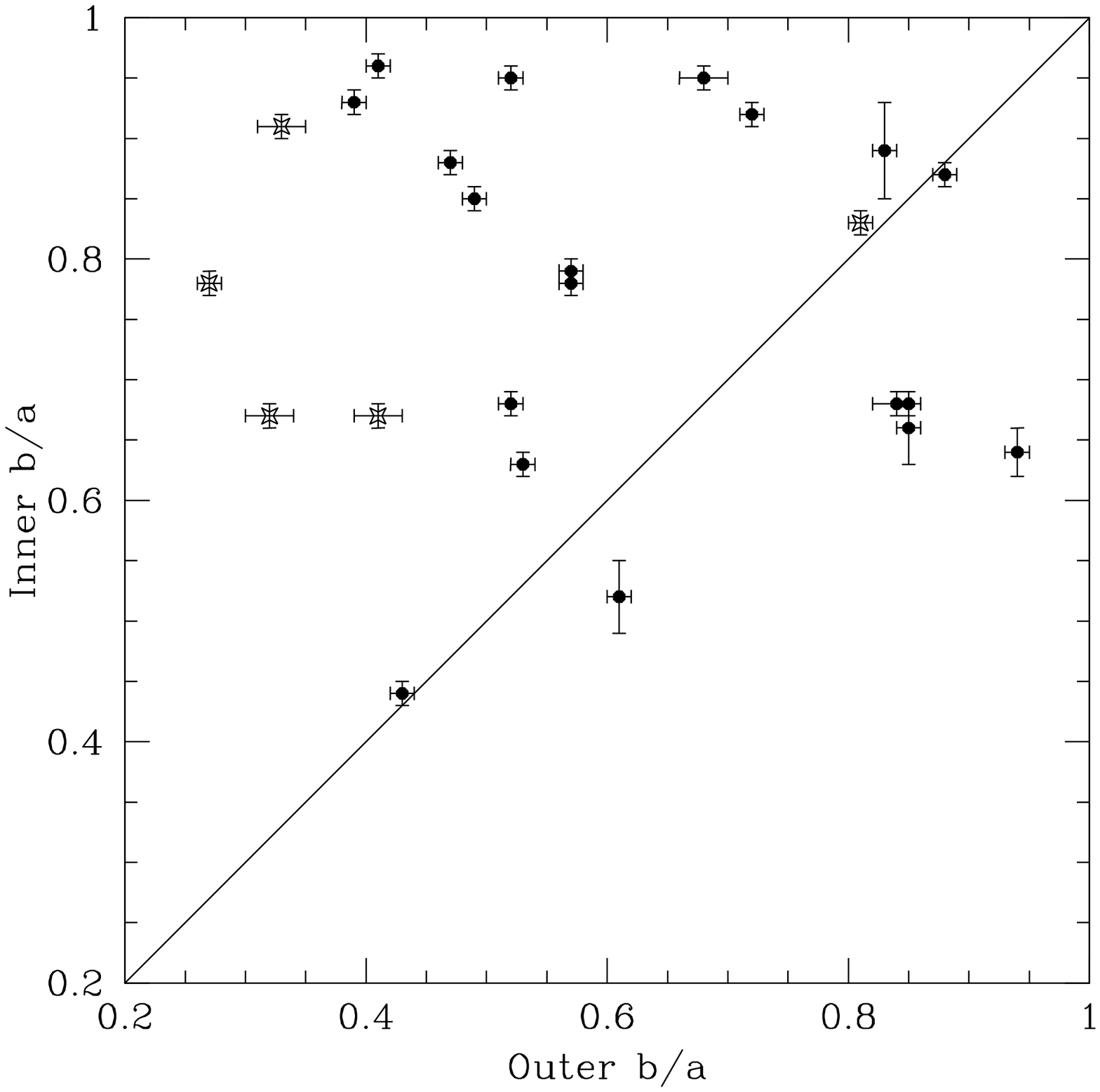}{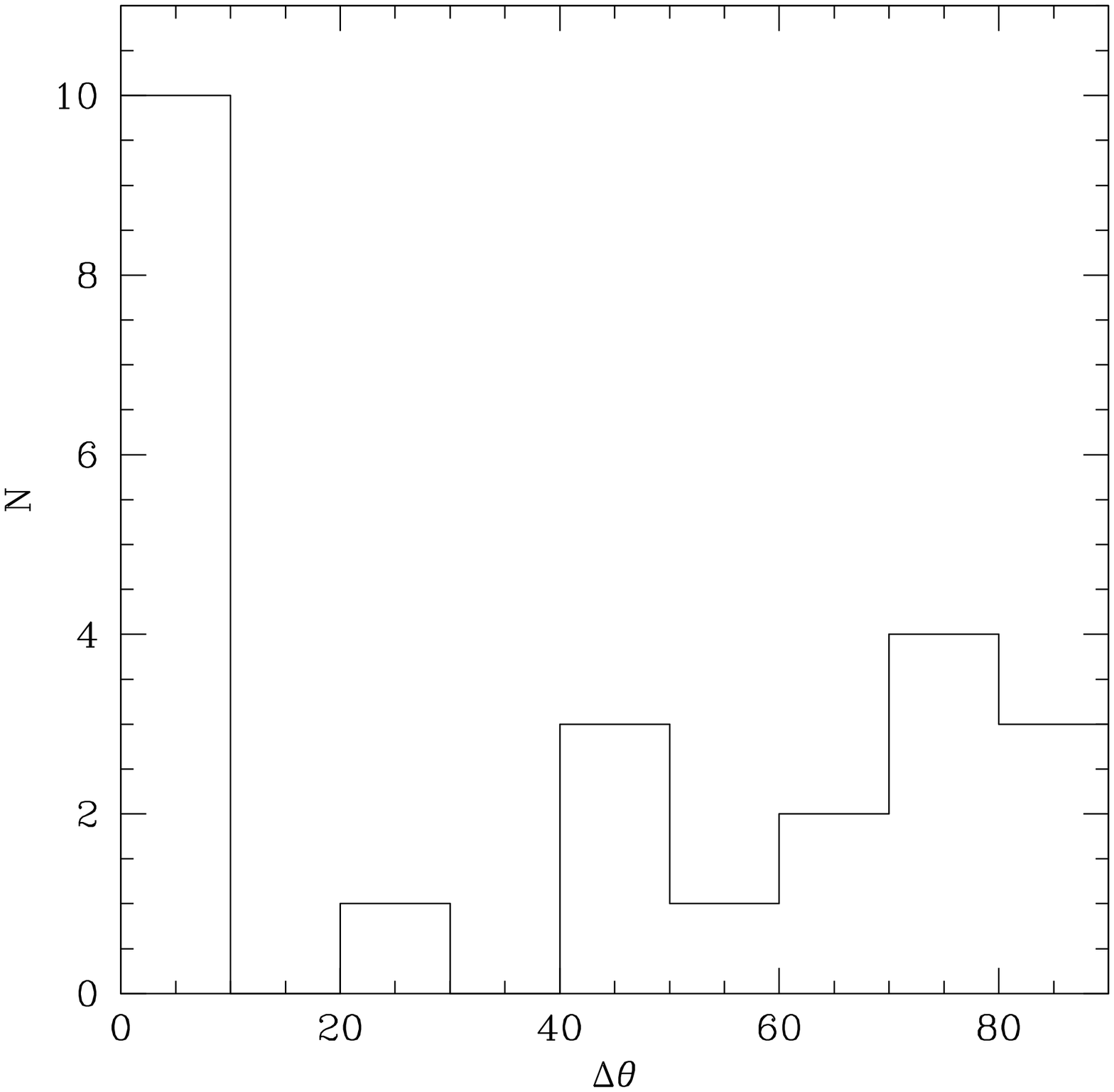}
\caption{
$(a)$ Axis ratios of the inner and outer components. The outer component is
typically flatter than the core. The symbols are the same as in Figure \ref{fig:r12}.
$(b)$ The distribution of position angle
offsets between the inner and outer components. The inner and outer components are aligned
to within 10$\degr$ in approximately 40\% of the BCGs.
}
\label{fig:paellcorr}
\end{figure*}

\begin{figure}
\plotone{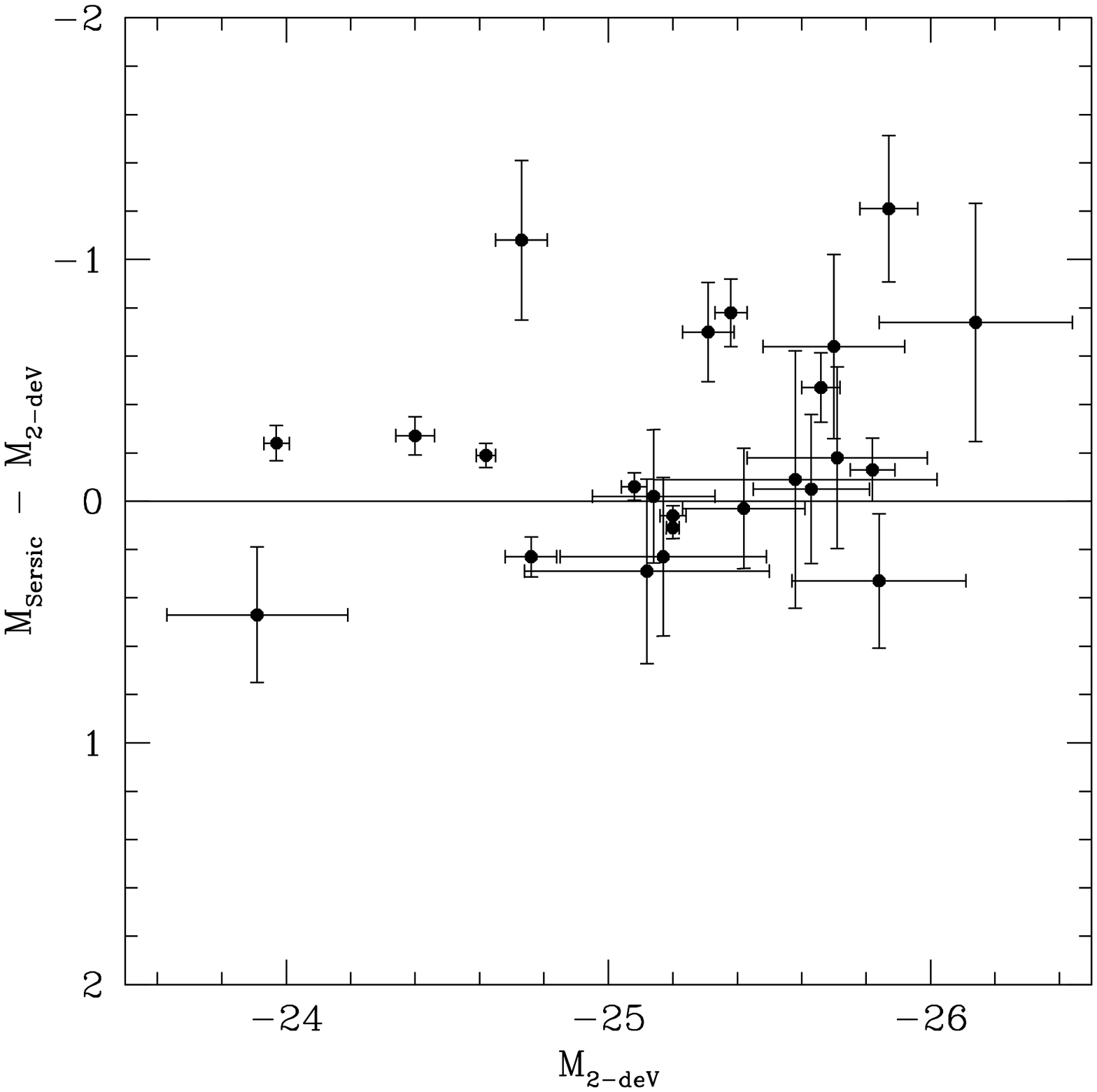}
\caption{Difference in absolute magnitudes between the \ser and \ddev
models. The magnitudes are comparable for 75\% of the galaxies, while
for the other 25\% the \ser magnitudes are $>$0.5 mag brighter.}
\label{fig:mdds}
\end{figure}

\begin{figure}
\plotone{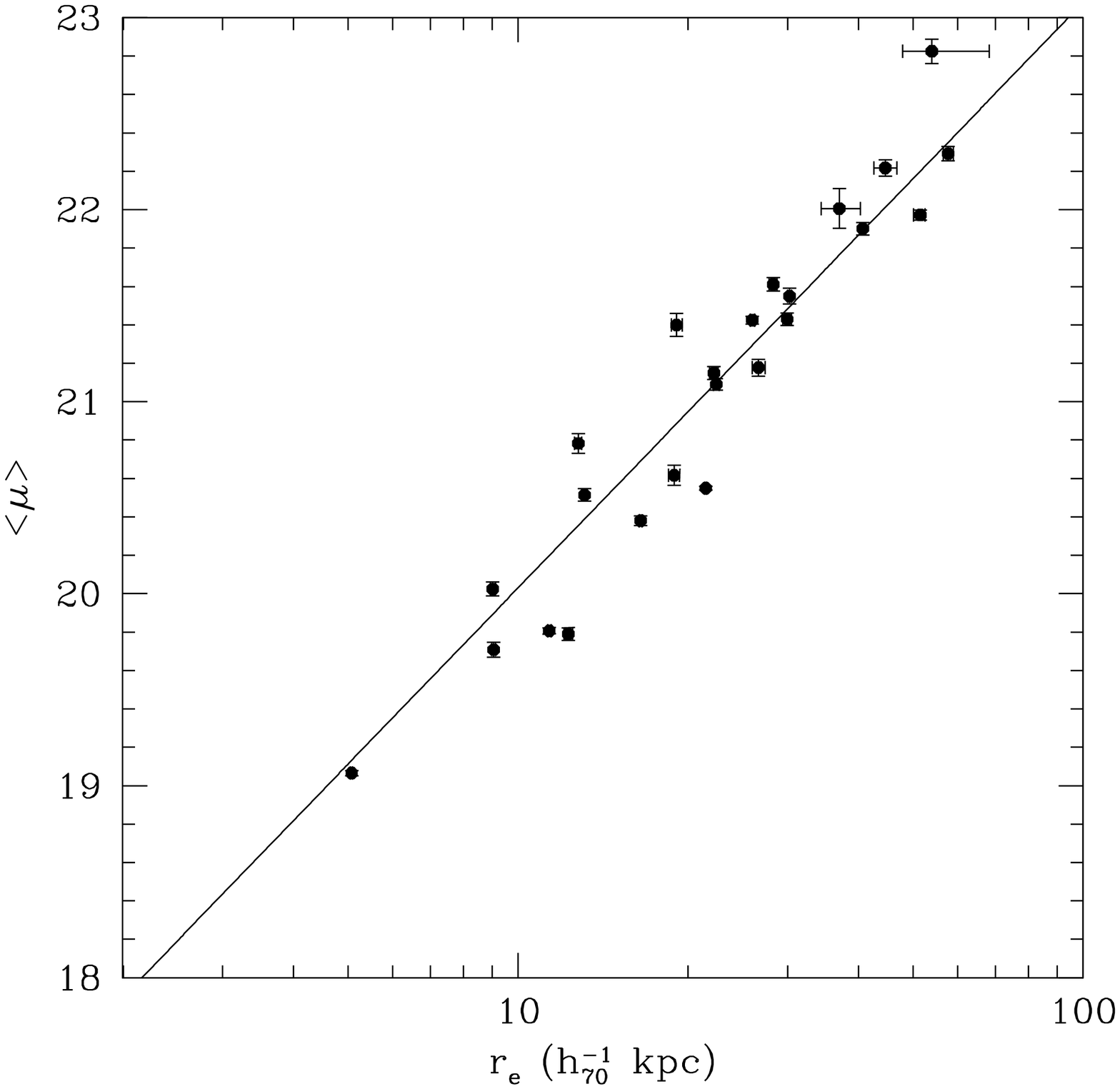}
\caption{Surface brightness vs. effective radius for the single \dev
parameterization. The slope of the solid line corresponds to the $<\mu>-r_e$
relation from \citet{jorgensen1996}, assuming constant velocity dispersion and
arbitrarily normalized to match the present data.  Despite the failure of the
\dev model at large radii, this parameterization yields a tight $<\mu>-r_e$
relation with a slope similar to the relation for normal elliptical galaxies.}
\label{fig:mueredev}
\end{figure}

\begin{figure*}
\plottwo{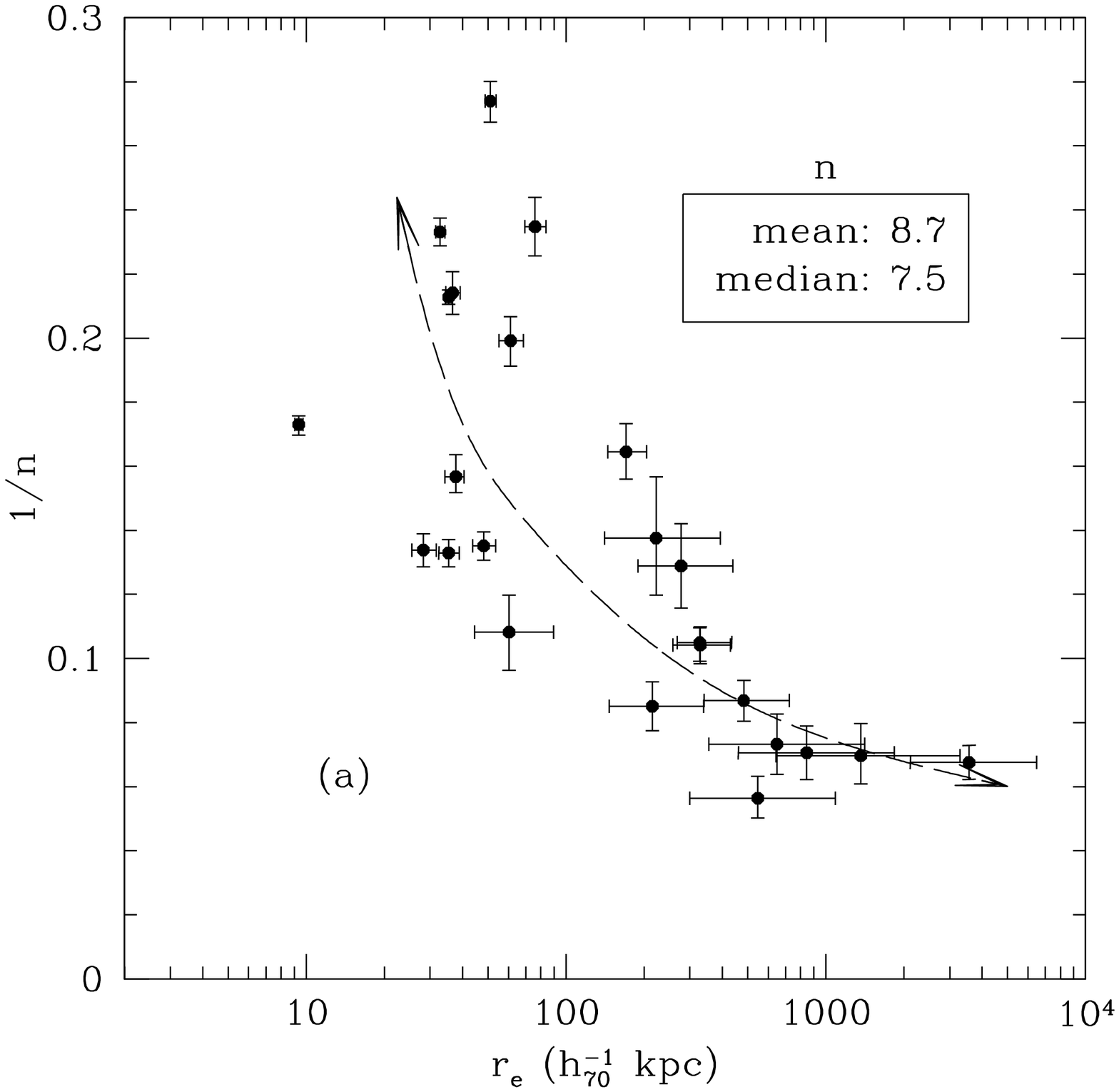}{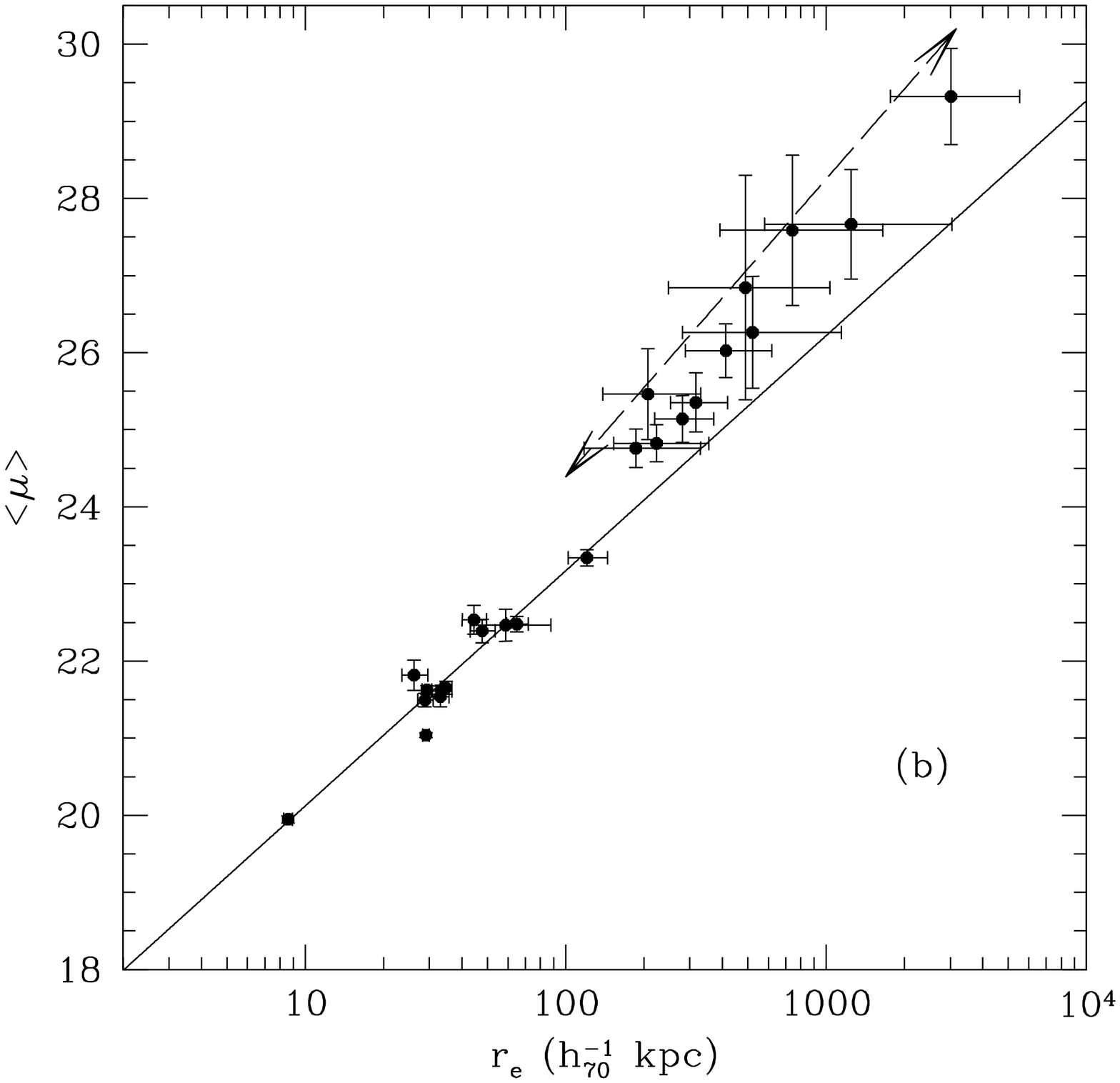}
\caption{
Parameters for the \ser model.  $(a)$ \ser index vs. effective radius. The
dashed arrow illustrates the intrinsic correlation between $r_e$ and $1/n$ in
the \ser model.  To generate this correlation vector we fit models to a single
galaxy multiple times with $n$ fixed to different values. The data indicate
that the true uncertainties in the \ser parameters are dominated by
degeneracies among the correlated parameters and are larger than the quoted
uncertainties.  Note that the length of the correlation vector is {\sl not} meant
to correspond to the size of the correlated uncertainties, but is simply set to
span the range
of observed $r_e$ values. $(b)$ Surface brightness vs. effective radius. The solid
line is the same as in Figure \ref{fig:mueredev}, while the dashed arrow again
indicates the correlation vector associated with changes in $n$. For large
effective radii the data lie parallel to this correlation vector. It is thus
difficult to decouple any true deviation from the FP relation from
correlated errors.}
\label{fig:sersiccorrelations}
\end{figure*}

\subsubsection{Centroid Offsets}
Finally, are the two components concentric? When modelling these galaxies with
GALFIT we allowed the centroids of the inner and outer components to vary
independently.  We find that the two components are concentric to within
1.2$\arcsec$ (less than the seeing FWHM) in all but one case. The exception is
Abell 1651, in which the centroids are offset by 9$\arcsec$ ($\sim 15$ \ho
kpc).

\subsection{Biases Arising from One-Component Models}  
\label{sec:correlations}

In preceding sections we discerned that the \ddev model provides a better
representation of the typical BCG surface brightness distribution than do \dev
or $r^{1/n}$ profiles, and quantified the properties of the components in this
model. We now return to the single component models, both to facilitate
comparison with previous work and to explore the relative biases that arise
from using these different models.

\subsubsection{Effects on Total Magnitude}

How does the choice of parametric model influence the total magnitude derived
for the brightest cluster galaxy?  The bulk of published BCG photometry
assumes either an \dev or $r^{1/n}$ profile to calculate a total
magnitude. The \dev model obviously underestimates the total luminosity
relative to the \ser and \ddev models, while Figure \ref{fig:profiles} argues
that in most cases the \ser and \ddev models should yield similar total
magnitudes.  Figure \ref{fig:mdds} demonstrates that this expectation is
correct for roughly three quarters of the galaxies. For the other 25\%, the
magnitudes derived from the \ser parameterization are more than half a
magnitude brighter (albeit typically with large uncertainties). In these
cases, the optimal \ser models have $n\ga9$ and thus contain significant
luminosity at $r>300$ \ho kpc (i.e. outside the fitting region).

Table
\ref{tab:global} lists the ensemble average and dispersion in the total
magnitudes derived using each parameterization. We also include in this table
the magnitudes within circular metric apertures of radii 300, 50, and 10 \ho
kpc.  Within the metric apertures the \ser and \ddev models yield consistent
magnitudes and dispersions, while the \dev model underestimates the flux by
$\sim30$\% within the 300 \ho kpc aperture. For the total magnitudes, the \ser
model yields significantly larger scatter, which is a consequence of the
extrapolation at $r>300$ \ho kpc for the galaxies with large $n$ described
above.  We note that the dispersions for the smallest apertures are consistent
with \citet{postman1995}, who found a dispersion of 0.327 mag within a metric
aperture of 10 $h_{100}^{-1}$ kpc.

\subsubsection{Effects on the Fundamental Plane}
\label{subsec:devsercor}

What is the effect of fitting a one-component model on the fundamental plane
derived for BCGs?  Figure \ref{fig:mueredev} shows the $<\mu> - r_e$ relation
for the single de Vaucouleurs model, which has a slope similar to that of the
$<\mu>-r_e$ projection of the fundamental plane (FP) for normal elliptical
galaxies \citep{jorgensen1996}.  It is interesting to note that the galaxies
roughly obey a fundamental plane type relation for the \dev fit despite the
demonstrated inferiority of this model.  Figure \ref{fig:sersiccorrelations}
shows the $1/n-r_e$ and $<\mu>-r_e$ relations for the \ser model. Because all
three parameters are strongly correlated, we explore the impact of changing
$n$ on the other parameters.  The additional degree of freedom provided by $n$
yields much larger effective radii (the largest $r_e$ in the \dev fits is $<
60$ \ho kpc) and fainter surface brightnesses (the faintest $<\mu>$ in the
\dev fits is $<$ 23). The slope of the $<\mu>-r_e$ relation is comparable to
the direction of the overlaid correlation vector at large $r_e$ (i.e. large
$n$), making any true deviation from the FP relation difficult to decouple
from correlated errors arising from uncertainty in $n$.  The scatter in the
$1/n-r_e$ plot is also predominantly along the direction of the correlation
vector. Together these plots illustrate a key limitation of the \ser model for
this analysis --- namely that systematic variations in $n$ can induce large
correlated uncertainties in the other parameters.

\section{Conclusions}
\label{sec:discussion}

Using highly uniform photometric data, we study the surface brightness
distribution of the brightest cluster galaxy (BCG) in each of 30 clusters at
$0.03<z<0.13$ that span a range of velocity dispersions and Bautz-Morgan
types.  In this first paper, we focus on the 24 clusters for which there is a
single, dominant BCG and no evidence of a substantial ongoing merger of
subclusters. We employ full two-dimensional fitting to model the surface
brightness profiles of these galaxies and their stellar halos within $r=300$
\ho kpc.  From this analysis, we reach several conclusions regarding the
nature of the diffuse optical light in clusters:

1.  The observed profiles are well described by a {\it two-component} model
consisting of two \dev profiles with different scale lengths, ellipticities,
and orientations. This model yields significantly lower $\chi^2$ than single
\dev or $r^{1/n}$ profiles and reproduces the position angle and ellipticity
gradients observed in these galaxies.  The nature of these gradients is such
that they are unlikely to originate from viewing a single component triaxial
system at random (\S \ref{subsec:twocomponents}). We conclude that the two
components should be considered physically distinct.

2.  The inner component is clearly associated with the BCG.  This component is
structurally similar to a normal bright elliptical galaxy.  The slope of the
$<\mu>- r_e$ projection of the fundamental plane is consistent with published
results over a comparable range of radii ($r_e\approx1-30$ \ho kpc), and the
observed axis ratios ($b/a=0.77\pm0.14$) are comparable to those observed in
normal ellipticals \citep{porter1991}.

3.  The outer component is consistent with arising from a population of
intracluster stars.  This intracluster light (ICL) component dominates the
luminosity, containing roughly 80-90\% of the total light of the two
components.  The large physical scale of the ICL component
($r_e\simeq20-500$\ho kpc) is similar to that expected for the dark matter
distribution \citep{dubinski1998} and is thus consistent with the picture in
which the intracluster stars trace the global cluster potential.  The ICL
component exhibits a tight $<\mu>-r_e$ relation, but with a steeper slope than
the corresponding projection of the elliptical galaxy fundamental plane ($<I>
\propto r_e^{-1.5}$ compared to $<I> \propto r_e^{-1.22}$, respectively).  The
ICL component is typically more elongated than either the inner BCG component
or normal ellipticals, with axis ratios comparable to those observed for the
distribution of cluster galaxies \citep{binggeli1982}.  The existence of a
distinct ICL component associated with the cluster potential is further
bolstered by kinematic data that detect rising velocity dispersions at
$r\ga10$ kpc in several systems (Abell 2029, \citet{dressler1979}; Abell 2199,
\citet{carter1999,kelson2002}), presumably due to the transition from the
bound stellar population of the BCG to the intracluster stellar population.

4.  The inner and outer components are strongly aligned
($|\Delta\theta|<10\degr$) in roughly 40\% of the clusters.  When they are not
aligned, the components tend toward high degrees of misalignment, suggesting
that accretion of infalling material may change the orientation of some BCGs
for a time.

5.  The ratio of scale-lengths for the outer and inner components,
$r_{e,outer}/r_{e,inner}$, is $\approx10-40$, with the distribution strongly
peaked at $r_{e,outer}/r_{e,inner}=35$.

Recent hydrodynamic simulations incorporate much of the relevant physics
necessary to produce populations of intracluster stars, and thus provide
quantitative predictions for the properties of the ICL.  These simulations
agree on several interesting points. Consistent with previous theoretical work
\citep[c.f.][]{richstone1983,miller1983}, the simulations predict that
intracluster stars produce roughly 20-50\% of the total cluster luminosity and
originate from tidal stripping \citep{larsen2004,murante2004,willman2004}.
The intracluster stars in these simulations form at higher median redshift
than stars that remain in cluster galaxies at $z=0$, concentrate more toward
the cluster center than the galaxies, and exceed the stellar mass density
within about half the virial radius
\citep{larsen2004,murante2004}. Additionally, formation of the ICL component
is found to be a continuing process, with significant growth in luminosity
since z=1 \citep{willman2004}.

We confirm that the ICL is a ubiquitous feature for clusters with a single,
dominant BCG, demonstrate that it dominates the combined luminosity of the
BCG+ICL components, and quantify some of the properties of the ICL. Recent
work by \citet{feldmeier2004} indicates that intracluster light is also
prevalent in clusters without dominant central galaxies, and thus together
these studies confirm that ICL is a generic feature of the cluster
environment. Our measurements of the extent of the ICL component and its
elongation, which is similar to published cluster galaxy spatial
distributions, further indicate that the evolution of the ICL is tied to the
cluster as a whole rather than to the BCG.

Finally, we emphasize that understanding the growth and accretion of brightest
cluster galaxies and the intracluster stellar populations is also of direct
relevance to studies of lower mass elliptical galaxies. \citet{graham1996}
interpreted the large \ser indices observed at $r\la25$ \ho kpc for their BCG
sample as the extension of an empirical correlation between $n$ and galaxy
mass observed by \citet{caon1993} for other elliptical galaxies, and
\citet{graham2001} have presented evidence for a correlation between $n$ and
the central black hole mass. We have now demonstrated that similarly large
\ser indices are obtained when BCGs are modelled to much larger radii.
Because the outer component in the \ddev model contributes a significant
fraction ($>$35\%) of the total luminosity within $r=25$ kpc for 75\% of the
BCGs in our sample, we argue that these large \ser indices are better
interpreted as arising from the dominance of this intracluster component at
large radii.  It is intriguing to consider whether the correlations observed
for lower mass systems are partly due to a correlation between galaxy mass and
the fractional luminosity existing in halos comprised of stars tidally
stripped from lower mass satellites.  We suspect that investigation along
these lines may provide a physical motivation for the observed empirical
correlations.

\section{Acknowledgements}
We are especially grateful for insightful conversations with Alister Graham
and Tod Lauer. We also thank Stefano Zibetti, Simon White, Heinz Andernach,
and the anonymous referee for helpful discussions and comments that improved
the paper.  AHG is supported by an NSF Astronomy and Astrophysics Postdoctoral
Fellowship under award AST-0407085.  AIZ is supported by NSF grant AST-0206084
and NASA LTSA grant NAG5-11108.  DZ acknowledges fellowships from the David
and Lucile Packard Foundation and the Alfred P. Sloan Foundation.  This
publication makes use of data products from the Two Micron All Sky Survey,
which is a joint project of the University of Massachusetts and the Infrared
Processing and Analysis Center/California Institute of Technology, funded by
the National Aeronautics and Space Administration and the National Science
Foundation.


\begin{appendix}
\label{sec:robustness}
\section{Robustness of Results}

We now explore whether either our ability to discriminate between models or
the resulting model parameters are compromised by systematic errors and
compare results for two clusters observed in multiple runs.

\subsection{Error in the sky level}
\label{sec:robustness:sky}
While GALFIT returns statistical uncertainties, we are undoubtedly limited by
{\it systematic} errors in deriving structural parameters. The most
significant source of such errors is the determined mean sky levels. In this
section we ask two questions: 1) does systematic error in the sky level
degrade our ability to discriminate between \ser and \ddev models, and 2) what
is the magnitude of the induced uncertainties in the structural parameters?

We estimated the $1\sigma$ uncertainty in the background sky levels in
\S \ref{sec:profiles:skylevel}. 
To assess whether this degree of uncertainty compromises our ability to
discriminate between models, we rerun GALFIT for all models with the sky level
artificially altered by $\pm1\sigma$. We find that the resulting $\chi^2$
differences between the $r^{1/n}$ and \ddev models are $\Delta\chi^2=1039$ and
$\Delta\chi^2=1089$ when we increase and decrease the sky, respectively, which
are consistent with our result from 
\S \ref{sec:results} ($\Delta\chi^2=1039$). 
We conclude that the statistical preference for \ddev models is not a result
of systematic error in the sky level.

From the same artifically altered data, we rederive the structural parameters.
The errors quoted in parentheses in Tables
\ref{tab:devparams}-\ref{tab:ddparams} correspond to the variations arising
from this change. For the \ser model the most sensitive parameters are $r_e$
and $n$, because a change in the outer structure can mimic a change in the sky
level. The derived effective radii and surface brightnesses are relatively
robust for cases with $n<10$, but are highly uncertain for larger indices (see
\S \ref{subsec:devsercor}).  For the \ddev model the parameters are remarkably
robust. The largest uncertainties arise for the outer component (for the same
reason as with the \ser model), and the uncertainty is proportional to the
effective radius of the outer component. Still, even for the cases where
$r_e\ga300$ \ho kpc (i.e. the size of the fitting region) the uncertainty in
$r_e$ is $\la 10$\%. These cases should obviously be viewed with caution (as
is always true when the fitting region is $<r_e$), but this robustness is
encouraging.

\subsection{Sensitivity to input PSF}
Another concern is whether the results are sensitive to the accuracy of the
input PSF. We assess this sensitivity by reanalyzing one of the BCGs (Abell
0122) using PSFs taken from other fields with seeing comparable to within
0.1$\arcsec$ (Abell 2376, Abell 3727, and Abell S1096).  Although drawn from
similar seeing conditions, these PSFs are structurally distinct.  The optimal
\ddev parameters are listed in Table \ref{tab:psfsensitivity} for each input
PSF.  The most sensitive parameter is $r_e$ for the inner component, which
increases up to 25\% when an incorrect PSF is used. We consider 25\% to be an
approximate upper bound on the systematic uncertainty in the inner $r_e$ due
to error in the input PSF. The outer component $r_e$, and $<\mu>$ for both
components, agree to within 7\% in all cases. Finally, $\chi^2$ changes by
$\la100$, which is significantly less than the differences between the three
functional forms tested in this paper.

A different concern is that the PSF convolved with the GALFIT models only
extends to $r=15\arcsec$. Could the inclusion of the PSF wings (Figure
\ref{fig:psf}) affect the results?  Convolution of the BCG profile with the
full PSF including the extended wings increases the surface brightness by
$\sim1$\% at $r\ga10\arcsec$, which should be small enough to not significantly
alter the derived structural parameters. To verify this assumption, we performed
modelling tests employing a large PSF that includes these wings. We find that the
most sensitive model parameter is the $r_e$ of the outer component in the \ddev
model, which typically decreases by $5-10$\% when the extended PSF is used. The
relative luminosities of the two components also change slightly, with the outer
component contributing $\sim2$\% less of the total luminosity.  In both cases this
bias due to using the smaller PSF corresponds to $\la1\sigma$ changes and thus
will not qualitatively alter any of the results described below.

\subsection{Sensitivity to masking}

Might the ``second component" observed at large radii be due to residual
emission from individual cluster galaxies?  If the masking is insufficient,
then flux from the outer regions of individual galaxies will appear as a
secondary halo in the surface brightness modelling.  This component would have
an ellipticity and orientation similar to the galaxy distribution, potentially
yielding gradients in both quantities as observed in our data.

We test this possibility by varying varying the degree of masking and
measuring the difference in the observed profiles.  We perform this test on
the BCG in Abell 2571, a galaxy for which the data indicates a significant
excess relative to \dev at $r\ga20$ \ho kpc.  We extend the object masks and
increase the fraction of masked pixels from 26\% to 53\%.  The best-fit
parameters change by less than the systematic uncertainties derived for
changes in the sky level, and the \ddev model maintains a significantly lower
$\chi^2$ than the other models ( $\chi^2_{deV}-\chi^2_{2-deV}=4300$ and
$\chi^2_{Sersic}-\chi^2_{2-deV}=1745$ ). The profiles derived using the normal
and enhanced masking are also consistent to within the 1$\sigma$ statistical
uncertainties for all radial bins.  We conclude that the observed outer
component is {\it not} an artifact arising from insufficient masking of
cluster galaxies.

\subsection{Clusters observed during multiple runs}
The two clusters in our sample that were observed in both 1999 and 2000, Abell
2721 and Abell S0296, provide an independent check on the impact of
systematics.  We model the BCGs independently for each data set and compare
the profiles and derived parameters for the BCGs.  For both galaxies the
azimuthally averaged surface brightness profiles are consistent to within the
statistical uncertainties at all radii larger than the seeing.  With only two
exceptions, the parameters derived from the two data sets are consistent to
within the systematic uncertainties for all models (Table
\ref{tab:devparams}---\ref{tab:ddparams}). The only exceptions are the
ellipticities for the \ser and \ddev models and the position angle shift,
$\Delta\theta$, in the \ddev model, which differ in some cases by as much
$\sim5\sigma$. We thus caution that the precision (typically $\sim 0.01$ in
ellipticity and $\sim$ few degrees in position angle) may be somewhat overestimated.

The most striking difference is in the $\Delta\chi^2$ values for Abell S0296,
which are significantly smaller for the 1999 data set. The likely origin of
this discrepancy is the poor data quality for the 1999 observations of this
cluster, which, as noted in Table \ref{tab:obs}, are compromised by
large-scale sky variations. Consequently, the comparison for this cluster can
be viewed as a worst-case test.  Nonetheless, both data sets yield the same
qualitative conclusions and comparable structural parameters, which again
argues that our results are robust against systematic errors.

\clearpage
\clearpage
\clearpage
\clearpage

\clearpage
\end{appendix}
\setcounter{table}{0}
\begin{deluxetable}{llcrccccl}
\footnotesize
\tabletypesize{\scriptsize}
\tablecaption{Cluster Sample}
\tablewidth{0pt}
\tablehead{
\colhead{}       & \colhead{} &\colhead{BM}   &\colhead{Date}  & \colhead{Exposure} & \colhead {}       & \colhead{      } & \colhead{2MASS}  & \colhead{} \\
\colhead{Cluster}& \colhead{$z_{BCG}$} &\colhead{type} &\colhead{of Run}& \colhead{time (s)}& \colhead {Airmass}& \colhead{FWHM($\arcsec$)}&\colhead{ID} & \colhead{Comments}
}
\startdata
Abell 0122  & .1127 &  I  & 10/2000 & 635 & 1.03 & 1.38 & 2MASX J00572288-2616528  & \\ 
Abell 0447  & .1123 &  I  & 10/2000 & 667 & 1.10 & 1.67 &     & \tablenotemark{a} \tablenotemark{b} \\
Abell 1651  & .0853 & I-II&  3/1995 & 285 & 1.12 & 1.47 & 2MASX J12592251-0411460  & \\ 
Abell 2376  & .0891 & I-II& 10/2000 & 673 & 1.19 & 1.33 & 2MASX J21460478-0927054  & \\ 
Abell 2400  & .0880 &  II & 10/2000 & 484 & 1.05 & 1.79 & 2MASX J21574250-1124402  & \\     
Abell 2401  & .0578 &  II & 10/2000 & 909 & 1.07 & 2.20 & 2MASX J21582246-2006145  & \\     
Abell 2405  & .0366 &     & 10/2000 & 698 & 1.13 & 1.34 & 2MASX J21594215-1748019  & \tablenotemark{c}\tablenotemark{d} \\ 
Abell 2571  & .1081 &  II & 11/1999 & 570 & 1.15 & 1.46 & 2MASX J23183367-0216356  & \\ 
Abell 2721  & .1149 &  II & 11/1999 & 694 & 1.01 & 1.79 & 2MASX J00055975-3443171    & \\     
            &       &     & 10/2000 & 808 & 1.03 & 1.33 &     & \\     
Abell 2730  & .120  &  II & 11-12/1999 & 818 & 1.08 & 2.38 & 2MASX J00095644-3541015 & \tablenotemark{e} \\     
Abell 2804  & .1121 & I-II& 12/1999 & 758 & 1.11 & 1.98 &  & \tablenotemark{b} \\ 
            &       &     & 10/2000 & 758 & 1.03 & 1.48 &     & \\ 
Abell 2811  & .1082 & I-II& 12/1999 & 432 & 1.30 & 2.10 & 2MASX J00420892-2832087    & \\     
Abell 2955  & .0945 &  II & 10/2000 & 695 & 1.14 & 1.41 & 2MASX J01570191-1701234    & \\     
Abell 2969  & .1271 &  I  & 12/1999 & 756 & 1.03 & 2.13 & 2MASX J02033533-4106002    & \\ 
Abell 2984  & .1044 &  I  & 10/2000 & 870 & 1.06 & 1.57 & 2MASX J02112484-4017261    & \\     
Abell 3094  & .0683 & I-II& 10/2000 & 745 & 1.02 & 1.60 & & \tablenotemark{b} \\ 
Abell 3112  & .0759 &  I  & 12/1999 & 530 & 1.23 & 1.87 & 2MASX J03175766-4414175     & \\     
Abell 3166  & .1171 &  I  & 10/2000 & 678 & 1.04 & 1.68 & 2MASX J03464387-3248454    & \\     
Abell 3194  & .0927 &  III& 10/2000 & 878 & 1.03 & 1.80 & & \tablenotemark{b} \\ 
Abell 3693  & .1237 &     & 10/2000 & 806 & 1.09 & 1.50 & 2MASX J20341991-3429387    & \tablenotemark{d}    \\ 
Abell 3705  & .0883 &  III& 10/2000 & 813 & 1.07 & 1.41 & 2MASX J20420438-3513067    & \\     
Abell 3727  & .1159 &  III& 10/2000 & 826 & 1.04 & 1.37 & 2MASX J20593652-3629085    & \\ 
Abell 3809  & .0616 &  III& 10/2000 & 790 & 1.14 & 1.37 & 2MASX J21465904-4353564    & \\ 
Abell 3920  & .1263 & I-II& 10/2000 & 878 & 1.23 & 2.30 & 2MASX J22492865-4053335    & \\     
Abell 4010  & .0963 & I-II& 11/1999 & 709 & 1.03 & 1.73 & 2MASX J23311412-3630405    & \\ 
APMC 020    & .1105 &     & 10/2000 & 808 & 1.05 & 1.42 & 2MASX J00133506-3443086 & \tablenotemark{d}    \\     
Abell S0084 & .1087 &  I  & 10/2000 & 763 & 1.07 & 1.25 & 2MASX J00492282-2931069    & \\ 
Abell S0296 & .0699 &  I  & 12/1999 & 514 & 1.33 & 2.30 & 2MASX J02463700-4222015 & \tablenotemark{f} \\ 
            &       &     & 10/2000 & 899 & 1.04 & 1.44 &     & \\     
Abell S0952 & .0898 &  II & 10/2000 & 813 & 1.16 & 1.41 & & \tablenotemark{b} \\ 
Abell S1096 & .1049 &  I-II & 10/2000 & 760 & 1.15 & 1.49 & & \tablenotemark{b} \\ 
\enddata

\tablenotetext{a}{The BCG in this cluster has a complicated core with multiple nuclei, which
are also visible in a {\sl Chandra} image of the cluster. }
\tablenotetext{b}{This BCG is not included in the statistical sample analyzed in this paper.}
\tablenotetext{c}{Abell 2405 is a superposition of two relatively poor systems at 11,000 and 27,000 km $s^{-1}$. 
Here we focus upon the lower redshift group.  }
\tablenotetext{d}{No published Bautz-Morgan type. Abell 2405 and Abell 3693 each have two discrete redshift peaks.
Published Bautz-Morgan types do exist for these clusters, but not for the redshift peaks that we are studying.}
\tablenotetext{e}{The listed value is the mean cluster redshift rather than the BCG redshift.}
\tablenotetext{f}{The 1999 data for Abell S0296 are poor. The images contain a large gradient not 
present in any of the other data.  }
%

\label{tab:obs}
\end{deluxetable}

\begin{deluxetable}{lrlrlrll}
\tabletypesize{\scriptsize}
\tablecaption{de Vaucouleurs Parameters}
\tablewidth{0pt}
\tablehead{
\colhead{Cluster} & \colhead{M} & \colhead{$\sigma_M$} & \colhead{$r_e$} & \colhead{$\sigma_{r_e}$} &
 \colhead{$b/a$} & \colhead{$\sigma_{b/a}$} & \colhead{$\chi^2_\nu$} 
}
\startdata
Abell 0122 &  $-24.98$ & $0.01\left(0.02\right)$ & 29.9 & $0.2\left(0.5\right)$ & 0.79 & 0.01 & 	1.123  \\
Abell 1651 &  $-25.44$ & $0.01\left(0.02\right)$ & 57.7 & $0.4\left(1.1\right)$ & 0.73 & 0.01 & 	0.985  \\
Abell 2376 &  $-23.69$ & $0.01\left(0.00\right)$ & 9.0 & $0.1\left(0.1\right)$ & 0.88 & 0.01 & 	0.987  \\
Abell 2400 &  $-24.77$ & $0.01\left(0.03\right)$ & 37.0 & $0.8\left(2.9\right)$ & 0.60 & 0.01 & 	0.997  \\
Abell 2401 &  $-24.28$ & $0.01\left(0.00\right)$ & 11.4 & $0.1\left(0.1\right)$ & 0.88 & 0.01 & 	1.043  \\
Abell 2405 &  $-23.18$ & $0.01\left(0.00\right)$ & 5.1 & $0.1\left(0.0\right)$ & 0.85 & 0.01 & 	1.002  \\
Abell 2571 &  $-24.66$ & $0.01\left(0.01\right)$ & 12.3 & $0.1\left(0.2\right)$ & 0.94 & 0.01 & 	1.046  \\
Abell 2721 (2000) &  $-25.19$ & $0.01\left(0.01\right)$ & 40.7 & $0.3\left(0.4\right)$ & 0.46 & 0.01 & 	1.020  \\
Abell 2721 (1999) &  $-25.18$ & $0.01\left(0.03\right)$ & 39.1 & $0.3\left(0.8\right)$ & 0.45 & 0.01 & 	1.005  \\
Abell 2730 &  $-25.01$ & $0.01\left(0.02\right)$ & 26.6 & $0.2\left(0.7\right)$ & 0.86 & 0.01 & 	1.066  \\
Abell 2811 &  $-24.78$ & $0.01\left(0.02\right)$ & 18.9 & $0.2\left(0.4\right)$ & 0.68 & 0.01 & 	1.022  \\
Abell 2955 &  $-24.54$ & $0.01\left(0.01\right)$ & 22.2 & $0.2\left(0.4\right)$ & 0.77 & 0.01 & 	1.068  \\
Abell 2969 &  $-24.92$ & $0.01\left(0.11\right)$ & 54.0 & $0.7\left(10.1\right)$ & 0.71 & 0.01 & 	1.022  \\
Abell 2984 &  $-24.65$ & $0.01\left(0.01\right)$ & 28.3 & $0.2\left(0.4\right)$ & 0.77 & 0.01 & 	1.109  \\
Abell 3112 &  $-25.47$ & $0.01\left(0.03\right)$ & 51.5 & $0.3\left(1.2\right)$ & 0.52 & 0.01 & 	0.987  \\
Abell 3166 &  $-24.05$ & $0.01\left(0.02\right)$ & 19.1 & $0.2\left(0.4\right)$ & 0.80 & 0.01 & 	1.014  \\
Abell 3693 &  $-24.15$ & $0.01\left(0.00\right)$ & 9.1 & $0.1\left(0.0\right)$ & 0.75 & 0.01 & 	1.033  \\
Abell 3705 &  $-24.01$ & $0.01\left(0.00\right)$ & 13.1 & $0.1\left(0.0\right)$ & 0.86 & 0.01 & 	1.041  \\
Abell 3727 &  $-23.79$ & $0.01\left(0.00\right)$ & 12.8 & $0.1\left(0.1\right)$ & 0.97 & 0.01 & 	1.036  \\
Abell 3809 &  $-24.47$ & $0.01\left(0.00\right)$ & 26.0 & $0.1\left(0.2\right)$ & 0.80 & 0.01 & 	1.006  \\
Abell 3920 &  $-24.79$ & $0.01\left(0.00\right)$ & 16.5 & $0.1\left(0.0\right)$ & 0.78 & 0.01 & 	0.994  \\
Abell 4010 &  $-25.00$ & $0.01\left(0.04\right)$ & 44.6 & $0.4\left(2.1\right)$ & 0.68 & 0.01 & 	1.019  \\
APMC 020 &  $-24.69$ & $0.01\left(0.00\right)$ & 22.4 & $0.1\left(0.2\right)$ & 0.62 & 0.01 & 	1.009  \\
Abell S0084 &  $-24.87$ & $0.01\left(0.00\right)$ & 30.2 & $0.2\left(0.0\right)$ & 0.93 & 0.01 & 	1.032  \\
Abell S0296 (2000) &  $-24.97$ & $0.01\left(0.00\right)$ & 21.5 & $0.1\left(0.1\right)$ & 0.69 & 0.01 & 	1.053  \\
Abell S0296 (1999) &  $-24.94$ & $0.01\left(0.01\right)$ & 21.6 & $0.1\left(0.3\right)$ & 0.65 & 0.01 & 	0.972  \\
\\
\enddata

\tablecomments{ The listed effective radius is $r_e\equiv\sqrt{ab}$ and is in
units \ho kpc. The absolute magnitudes, which are total magnitudes that include extrapolation beyond $r=300$ \ho kpc, are {\sl not} corrected for extinction or k-dimming. Both corrections are typically $<$0.1 mag for galaxies in our sample. The 
column $b/a$ gives the ratio of the minor to major axes. In the last
column $\chi^2_\nu$ is the reduced $\chi^2$. For clusters observed during multiple years, the ensemble $\chi^2_\nu$ value includes only the 2000 data.
Parameter uncertainties corresponding to the $1\sigma$ sky level
uncertainties are given in parentheses.}
\label{tab:devparams}
\end{deluxetable}

\begin{deluxetable}{lrlrcrlrlr}
\tabletypesize{\scriptsize}
\tablecaption{Sersic Parameters}
\tablewidth{0pt}
\tablehead{
\colhead{Cluster} & \colhead{M} & \colhead{$\sigma_M$} &  \colhead{$r_e$} & \colhead{$\sigma_{r_e}$} &
 \colhead{$n$} & \colhead{$\sigma_n$} &  \colhead{$b/a$} & \colhead{$\sigma_{b/a}$} & \colhead{$\Delta\chi^2_d$} 
}
\startdata
Abell 0122 & $-26.13$ &0.03 (0.13) & 280.7 & 16.9 $(^{+86.7}_{-59.9} )$ & 9.60 & 0.14 $(^{+0.57}_{-0.50} )$ & 0.73 & 0.01 & 	$-6032$  \\
Abell 1651 & $-25.51$ &0.01 (0.07) & 65.0 & 1.3 $(^{+6.7}_{-5.6} )$ & 4.26 & 0.04 $(^{+0.17}_{-0.16} )$ & 0.73 & 0.01 & 	$-38$  \\
Abell 2376 & $-24.21$ &0.02 (0.06) & 26.2 & 1.0 $(^{+3.2}_{-2.5} )$ & 7.47 & 0.11 $(^{+0.31}_{-0.27} )$ & 0.86 & 0.01 & 	$-2073$  \\
Abell 2400 & $-24.94$ &0.02 (0.07) & 47.8 & 1.4 $(^{+5.7}_{-4.6} )$ & 5.02 & 0.06 $(^{+0.21}_{-0.18} )$ & 0.61 & 0.01 & 	$-195$  \\
Abell 2401 & $-24.81$ &0.01 (0.04) & 33.3 & 0.5 $(^{+3.3}_{-2.7} )$ & 7.52 & 0.04 $(^{+0.26}_{-0.23} )$ & 0.89 & 0.01 & 	$-14411$  \\
Abell 2405 & $-23.44$ &0.01 (0.02) & 8.6 & 0.1 $(^{+0.3}_{-0.3} )$ & 5.78 & 0.03 $(^{+0.11}_{-0.09} )$ & 0.84 & 0.01 & 	$-5774$  \\
Abell 2571 & $-25.39$ &0.02 (0.16) & 58.8 & 2.4 $(^{+28.6}_{-15.6} )$ & 9.24 & 0.13 $(^{+1.14}_{-0.89} )$ & 0.95 & 0.01 & 	$-4580$  \\
Abell 2721 (2000) & $-25.09$ &0.01 (0.04) & 34.7 & 0.6 $(^{+1.8}_{-1.6} )$ & 3.65 & 0.03 $(^{+0.09}_{-0.08} )$ & 0.46 & 0.01 & 	$-84$  \\
Abell 2721 (1999) & $-25.12$ &0.01 (0.07) & 35.5 & 0.7 $(^{+4.3}_{-3.3} )$ & 3.77 & 0.04 $(^{+0.22}_{-0.18} )$ & 0.45 & 0.01 & 	$-19$  \\
Abell 2730 & $-26.88$ &0.06 (0.39) & 1250.0 & 177.4 $(^{+1759.3}_{-661.3} )$ & 14.36 & 0.37 $(^{+2.09}_{-1.81} )$ & 0.84 & 0.01 & 	$-5253$  \\
Abell 2811 & $-26.34$ &0.06 (0.31) & 522.0 & 75.4 $(^{+607.3}_{-235.7} )$ & 13.65 & 0.41 $(^{+2.01}_{-1.55} )$ & 0.65 & 0.01 & 	$-3351$  \\
Abell 2955 & $-26.01$ &0.03 (0.19) & 412.5 & 28.8 $(^{+202.1}_{-122.2} )$ & 11.52 & 0.17 $(^{+0.91}_{-0.79} )$ & 0.73 & 0.01 & 	$-9107$  \\
Abell 2969 & $-25.67$ &0.03 (0.30) & 185.9 & 9.3 $(^{+142.8}_{-68.1} )$ & 7.27 & 0.12 $(^{+1.08}_{-0.89} )$ & 0.70 & 0.01 & 	$-390$  \\
Abell 2984 & $-27.08$ &0.06 (0.29) & 3019.2 & 374.7 $(^{+2451.5}_{-1222.5} )$ & 14.81 & 0.28 $(^{+1.25}_{-1.09} )$ & 0.72 & 0.01 & 	$-10569$  \\
Abell 3112 & $-25.95$ &0.01 (0.11) & 120.5 & 2.5 $(^{+24.0}_{-18.2} )$ & 6.08 & 0.04 $(^{+0.33}_{-0.31} )$ & 0.50 & 0.01 & 	$-3406$  \\
Abell 3166 & $-25.81$ &0.08 (0.32) & 741.5 & 143.8 $(^{+871.5}_{-337.0} )$ & 14.16 & 0.52 $(^{+1.93}_{-1.50} )$ & 0.77 & 0.01 & 	$-2693$  \\
Abell 3693 & $-25.68$ &0.10 (0.25) & 489.7 & 142.3 $(^{+483.2}_{-223.7} )$ & 17.74 & 1.03 $(^{+2.15}_{-1.92} )$ & 0.80 & 0.01 & 	$-2535$  \\
Abell 3705 & $-24.64$ &0.02 (0.05) & 44.4 & 1.7 $(^{+4.9}_{-4.1} )$ & 7.40 & 0.10 $(^{+0.25}_{-0.23} )$ & 0.85 & 0.01 & 	$-3419$  \\
Abell 3727 & $-25.16$ &0.05 (0.20) & 207.1 & 24.4 $(^{+117.8}_{-65.8} )$ & 11.75 & 0.32 $(^{+1.15}_{-0.96} )$ & 0.93 & 0.01 & 	$-2494$  \\
Abell 3809 & $-24.53$ &0.01 (0.02) & 29.3 & 0.3 $(^{+1.2}_{-1.1} )$ & 4.29 & 0.02 $(^{+0.08}_{-0.08} )$ & 0.80 & 0.01 & 	$-165$  \\
Abell 3920 & $-25.14$ &0.01 (0.04) & 33.0 & 0.9 $(^{+2.5}_{-3.0} )$ & 6.38 & 0.08 $(^{+0.21}_{-0.27} )$ & 0.77 & 0.01 & 	$-1296$  \\
Abell 4010 & $-25.89$ &0.03 (0.25) & 223.4 & 10.6 $(^{+131.0}_{-69.8} )$ & 7.76 & 0.10 $(^{+0.88}_{-0.72} )$ & 0.65 & 0.01 & 	$-3198$  \\
APMC 020 & $-24.83$ &0.01 (0.04) & 28.8 & 0.5 $(^{+2.0}_{-1.7} )$ & 4.67 & 0.04 $(^{+0.15}_{-0.14} )$ & 0.62 & 0.01 & 	$-270$  \\
Abell S0084 & $-26.16$ &0.04 (0.13) & 315.8 & 24.1 $(^{+98.5}_{-60.2} )$ & 9.53 & 0.17 $(^{+0.56}_{-0.41} )$ & 0.93 & 0.01 & 	$-3605$  \\
Abell S0296 (2000) & $-25.14$ &0.01 (0.01) & 29.1 & 0.2 $(^{+0.9}_{-0.8} )$ & 4.70 & 0.01 $(^{+0.05}_{-0.05} )$ & 0.68 & 0.01 & 	$-3338$  \\
Abell S0296 (1999) & $-25.04$ &0.01 (0.04) & 25.9 & 0.3 $(^{+1.8}_{-1.5} )$ & 4.54 & 0.03 $(^{+0.15}_{-0.15} )$ & 0.65 & 0.01 & 	$-302$  \\
\\
{\bf Average} & & & & & & & & & $-3680$   \\
\enddata

\tablecomments{ In the last column $\Delta\chi^2_d\equiv\chi^2_{sersic}-\chi^2_{de Vaucouleurs}$.
For clusters observed during multiple years, the average and ensemble values are computed using only the 2000 data.
Parameter uncertainties corresponding to the $1\sigma$ sky level
uncertainties are given in parentheses. As in Table \ref{tab:devparams}, the absolute magnitudes are not corrected for
extinction or k-dimming. These absolute magnitudes are total magnitudes that include extrapolation beyond $r=300$ \ho kpc.}

\label{tab:serparams}
\end{deluxetable}

\begin{deluxetable}{lrlrcrlrcrlrr}
\tabletypesize{\scriptsize}
\tablecaption{\ddev Parameters}
\tablewidth{0pt}
\tablehead{
\colhead{Cluster} & \colhead{$M_{total}$} & \colhead{$\sigma_{M_{total}}$} & \colhead{$|\Delta\theta|$} & \colhead{$\sigma_{|\Delta\theta|}$} &
                    \colhead{$M$} & \colhead{$\sigma_{M}$} & \colhead{$r_{e}$} & \colhead{$\sigma_{r_e}$} &
	            \colhead{$b/a$} & \colhead{$\sigma_{b/a}$}  &\colhead{$\Delta\chi^2_d$} & \colhead{$\Delta\chi^2_s$}
}
\startdata
Abell 0122 & $-25.66$ & $(^{+0.06}_{-0.06})$ & 76.5 & 1.1 (0.5) & $-25.56$ & 0.01 $(^{+0.06}_{-0.06})$ & 107.9 & 2.9 $(^{+9.3}_{-7.4})$ & 0.53 & 0.01 & $-9189$ & $-3156$ \\ 
 &      &                    &      &              & $-22.97$ & 0.03 $(^{+0.05}_{-0.05})$ & 4.7 & 0.1 $(^{+0.2}_{-0.2})$ & 0.63 & 0.01 &    &    \\ 
Abell 1651 & $-25.84$ & $(^{+0.17}_{-0.27})$ & 9.2 & 1.3 (0.5) & $-24.95$ & 0.07 $(^{+0.48}_{-0.53})$ & 477.2 & 53.1 $(^{+239.1}_{-130.5})$ & 0.27 & 0.01 & $-328$ & $-289$ \\ 
 &      &                    &      &              & $-25.21$ & 0.01 $(^{+0.00}_{-0.02})$ & 48.2 & 0.5 $(^{+0.6}_{-0.2})$ & 0.78 & 0.01 &    &    \\ 
Abell 2376 & $-23.97$ & $(^{+0.03}_{-0.04})$ & 68.5 & 13.2 (5.5) & $-23.87$ & 0.01 $(^{+0.02}_{-0.02})$ & 18.4 & 0.6 $(^{+2.0}_{-1.0})$ & 0.83 & 0.01 & $-2197$ & $-123$ \\ 
 &      &                    &      &              & $-21.33$ & 0.08 $(^{+0.09}_{-0.20})$ & 1.0 & 0.1 $(^{+0.2}_{-0.1})$ & 0.89 & 0.04 &    &    \\ 
Abell 2400 & $-25.17$ & $(^{+0.18}_{-0.32})$ & 10.0 & 1.5 (2.5) & $-24.61$ & 0.04 $(^{+0.19}_{-0.44})$ & 184.6 & 44.9 $(^{+169.7}_{-75.2})$ & 0.41 & 0.02 & $-339$ & $-144$ \\ 
 &      &                    &      &              & $-24.19$ & 0.08 $(^{+0.17}_{-0.11})$ & 20.4 & 1.0 $(^{+0.8}_{-1.4})$ & 0.67 & 0.01 &    &    \\ 
Abell 2401 & $-24.62$ & $(^{+0.03}_{-0.03})$ & 24.4 & 3.2 (0.4) & $-24.48$ & 0.01 $(^{+0.02}_{-0.02})$ & 27.6 & 0.4 $(^{+2.3}_{-1.9})$ & 0.88 & 0.01 & $-15350$ & $-939$ \\ 
 &      &                    &      &              & $-22.30$ & 0.03 $(^{+0.10}_{-0.12})$ & 1.8 & 0.1 $(^{+0.2}_{-0.2})$ & 0.87 & 0.01 &    &    \\ 
Abell 2405 & $-23.91$ & $(^{+0.17}_{-0.28})$ & 43.2 & 2.5 (0.3) & $-23.38$ & 0.02 $(^{+0.25}_{-0.40})$ & 93.9 & 4.7 $(^{+67.3}_{-33.5})$ & 0.81 & 0.01 & $-7687$ & $-1912$ \\ 
 &      &                    &      &              & $-22.89$ & 0.01 $(^{+0.06}_{-0.05})$ & 3.8 & 0.1 $(^{+0.2}_{-0.2})$ & 0.83 & 0.01 &    &    \\ 
Abell 2571 & $-25.42$ & $(^{+0.15}_{-0.19})$ & 83.6 & 0.8 (0.5) & $-25.11$ & 0.01 $(^{+0.17}_{-0.22})$ & 107.7 & 5.0 $(^{+43.5}_{-27.9})$ & 0.52 & 0.01 & $-7247$ & $-2667$ \\ 
 &      &                    &      &              & $-23.89$ & 0.02 $(^{+0.08}_{-0.09})$ & 5.4 & 0.1 $(^{+0.4}_{-0.4})$ & 0.68 & 0.01 &    &    \\ 
Abell 2721 (2000) & $-25.20$ & $(^{+0.02}_{-0.02})$ & 1.1 & 1.3 (0.4) & $-25.02$ & 0.03 $(^{+0.04}_{-0.04})$ & 47.9 & 0.8 $(^{+1.8}_{-1.3})$ & 0.43 & 0.01 & $-810$ & $-725$ \\ 
 &      &                    &      &              & $-23.14$ & 0.14 $(^{+0.08}_{-0.08})$ & 18.1 & 1.6 $(^{+1.2}_{-1.1})$ & 0.44 & 0.01 &    &    \\ 
Abell 2721 (1999) & $-25.19$ & $(^{+0.03}_{-0.04})$ & 9.7 & 1.4 (0.5) & $-25.04$ & 0.04 $(^{+0.06}_{-0.06})$ & 45.8 & 0.9 $(^{+2.7}_{-2.1})$ & 0.43 & 0.01 & $-391$ & $-372$ \\ 
 &      &                    &      &              & $-22.97$ & 0.21 $(^{+0.09}_{-0.14})$ & 16.1 & 1.9 $(^{+2.0}_{-1.8})$ & 0.42 & 0.02 &    &    \\ 
Abell 2730 & $-26.14$ & $(^{+0.24}_{-0.30})$ & 40.3 & 5.0 (4.6) & $-26.00$ & 0.03 $(^{+0.25}_{-0.32})$ & 245.9 & 13.5 $(^{+121.9}_{-78.4})$ & 0.72 & 0.01 & $-5382$ & $-129$ \\ 
 &      &                    &      &              & $-23.85$ & 0.03 $(^{+0.19}_{-0.16})$ & 8.1 & 0.2 $(^{+1.2}_{-1.3})$ & 0.92 & 0.01 &    &    \\ 
Abell 2811 & $-25.70$ & $(^{+0.22}_{-0.22})$ & 8.7 & 3.3 (4.9) & $-25.53$ & 0.03 $(^{+0.22}_{-0.23})$ & 139.5 & 8.6 $(^{+59.7}_{-44.4})$ & 0.49 & 0.01 & $-3556$ & $-204$ \\ 
 &      &                    &      &              & $-23.59$ & 0.03 $(^{+0.22}_{-0.15})$ & 5.5 & 0.2 $(^{+0.8}_{-1.0})$ & 0.85 & 0.01 &    &    \\ 
Abell 2955 & $-25.31$ & $(^{+0.08}_{-0.08})$ & 70.0 & 1.9 (2.2) & $-25.22$ & 0.01 $(^{+0.08}_{-0.08})$ & 93.5 & 2.3 $(^{+11.3}_{-9.8})$ & 0.57 & 0.01 & $-10927$ & $-1820$ \\ 
 &      &                    &      &              & $-22.61$ & 0.03 $(^{+0.09}_{-0.08})$ & 3.4 & 0.1 $(^{+0.3}_{-0.3})$ & 0.79 & 0.01 &    &    \\ 
Abell 2969 & $-25.58$ & $(^{+0.26}_{-0.44})$ & 55.5 & 9.4 (28.9) & $-25.35$ & 0.03 $(^{+0.27}_{-0.49})$ & 232.0 & 26.7 $(^{+203.6}_{-83.2})$ & 0.39 & 0.01 & $-841$ & $-450$ \\ 
 &      &                    &      &              & $-23.80$ & 0.08 $(^{+0.21}_{-0.18})$ & 19.7 & 0.8 $(^{+1.8}_{-1.7})$ & 0.93 & 0.01 &    &    \\ 
Abell 2984 & $-25.87$ & $(^{+0.01}_{-0.09})$ & 64.5 & 8.0 (5.0) & $-25.79$ & 0.01 $(^{+0.01}_{-0.09})$ & 202.7 & 5.4 $(^{+23.2}_{-2.4})$ & 0.52 & 0.01 & $-11873$ & $-1303$ \\ 
 &      &                    &      &              & $-23.04$ & 0.02 $(^{+0.01}_{-0.04})$ & 6.1 & 0.1 $(^{+0.2}_{-0.0})$ & 0.95 & 0.01 &    &    \\ 
Abell 3112 & $-25.82$ & $(^{+0.07}_{-0.07})$ & 42.3 & 15.5 (29.5) & $-25.74$ & 0.01 $(^{+0.06}_{-0.06})$ & 102.1 & 1.7 $(^{+12.4}_{-9.4})$ & 0.41 & 0.01 & $-5666$ & $-2260$ \\ 
 &      &                    &      &              & $-22.95$ & 0.04 $(^{+0.16}_{-0.17})$ & 8.3 & 0.2 $(^{+0.9}_{-0.7})$ & 0.96 & 0.01 &    &    \\ 
Abell 3166 & $-24.73$ & $(^{+0.07}_{-0.08})$ & 82.4 & 2.1 (0.1) & $-24.65$ & 0.02 $(^{+0.07}_{-0.08})$ & 69.2 & 2.6 $(^{+8.6}_{-6.4})$ & 0.61 & 0.01 & $-3355$ & $-662$ \\ 
 &      &                    &      &              & $-21.92$ & 0.04 $(^{+0.08}_{-0.08})$ & 1.7 & 0.1 $(^{+0.2}_{-0.1})$ & 0.52 & 0.03 &    &    \\ 
Abell 3693 & $-25.63$ & $(^{+0.18}_{-0.18})$ & 71.5 & 5.4 (0.6) & $-25.46$ & 0.05 $(^{+0.20}_{-0.21})$ & 314.5 & 24.4 $(^{+61.5}_{-65.4})$ & 0.84 & 0.02 & $-3713$ & $-1178$ \\ 
 &      &                    &      &              & $-23.56$ & 0.02 $(^{+0.10}_{--0.04})$ & 4.9 & 0.1 $(^{+-0.3}_{-0.5})$ & 0.68 & 0.01 &    &    \\ 
Abell 3705 & $-24.38$ & $(^{+0.04}_{-0.08})$ & 2.5 & 4.1 (2.6) & $-24.27$ & 0.01 $(^{+0.03}_{-0.00})$ & 30.8 & 1.0 $(^{+10.9}_{-2.6})$ & 0.93 & 0.01 & $-3522$ & $-103$ \\ 
 &      &                    &      &              & $-21.85$ & 0.07 $(^{+0.18}_{-0.62})$ & 2.1 & 0.1 $(^{+1.5}_{-0.3})$ & 0.62 & 0.02 &    &    \\ 
Abell 3727 & $-25.14$ & $(^{+0.17}_{-0.19})$ & 88.3 & 7.7 (1.5) & $-24.97$ & 0.05 $(^{+0.19}_{-0.21})$ & 215.2 & 17.1 $(^{+54.6}_{-43.7})$ & 0.68 & 0.02 & $-2953$ & $-458$ \\ 
 &      &                    &      &              & $-23.07$ & 0.03 $(^{+0.05}_{-0.04})$ & 6.1 & 0.1 $(^{+0.2}_{-0.3})$ & 0.95 & 0.01 &    &    \\ 
Abell 3809 & $-24.76$ & $(^{+0.07}_{-0.08})$ & 4.2 & 1.4 (0.6) & $-23.85$ & 0.03 $(^{+0.13}_{-0.16})$ & 126.8 & 17.2 $(^{+31.4}_{-22.0})$ & 0.33 & 0.02 & $-1096$ & $-931$ \\ 
 &      &                    &      &              & $-24.14$ & 0.03 $(^{+0.02}_{-0.02})$ & 20.8 & 0.4 $(^{+0.1}_{-0.2})$ & 0.91 & 0.01 &    &    \\ 
Abell 3920 & $-25.08$ & $(^{+0.02}_{-0.04})$ & 0.5 & 3.3 (0.4) & $-24.80$ & 0.01 $(^{+0.01}_{-0.04})$ & 45.6 & 3.2 $(^{+4.4}_{-3.2})$ & 0.85 & 0.01 & $-1421$ & $-124$ \\ 
 &      &                    &      &              & $-23.48$ & 0.10 $(^{+0.07}_{-0.05})$ & 5.4 & 0.4 $(^{+0.1}_{-0.2})$ & 0.68 & 0.01 &    &    \\ 
Abell 4010 & $-25.71$ & $(^{+0.21}_{-0.28})$ & 9.4 & 3.2 (7.4) & $-25.56$ & 0.02 $(^{+0.19}_{-0.28})$ & 186.3 & 10.7 $(^{+107.4}_{-58.1})$ & 0.47 & 0.01 & $-3579$ & $-381$ \\ 
 &      &                    &      &              & $-23.52$ & 0.06 $(^{+0.37}_{-0.28})$ & 13.7 & 0.6 $(^{+2.8}_{-3.0})$ & 0.88 & 0.01 &    &    \\ 
APMC 020 & $-25.12$ & $(^{+0.15}_{-0.38})$ & 2.4 & 1.0 (0.3) & $-24.34$ & 0.04 $(^{+0.25}_{-0.62})$ & 180.2 & 34.4 $(^{+259.1}_{-71.3})$ & 0.32 & 0.02 & $-737$ & $-467$ \\ 
 &      &                    &      &              & $-24.39$ & 0.03 $(^{+0.07}_{-0.09})$ & 17.6 & 0.4 $(^{+0.9}_{-0.7})$ & 0.67 & 0.01 &    &    \\ 
Abell S0084 & $-25.38$ & $(^{+0.03}_{-0.05})$ & 73.6 & 3.0 (1.3) & $-25.33$ & 0.01 $(^{+0.03}_{-0.05})$ & 72.0 & 1.6 $(^{+5.4}_{-3.9})$ & 0.85 & 0.01 & $-4027$ & $-422$ \\ 
 &      &                    &      &              & $-21.94$ & 0.04 $(^{+0.08}_{-0.10})$ & 2.0 & 0.1 $(^{+0.2}_{-0.2})$ & 0.66 & 0.03 &    &    \\ 
Abell S0296 (2000) & $-25.20$ & $(^{+0.02}_{-0.04})$ & 7.1 & 0.8 (0.9) & $-24.89$ & 0.01 $(^{+0.02}_{-0.00})$ & 46.1 & 1.2 $(^{+7.5}_{-3.1})$ & 0.57 & 0.01 & $-7429$ & $-4091$ \\ 
 &      &                    &      &              & $-23.67$ & 0.05 $(^{+0.10}_{-0.19})$ & 10.0 & 0.3 $(^{+0.9}_{-0.3})$ & 0.78 & 0.01 &    &    \\ 
Abell S0296 (1999) & $-25.05$ & $(^{+0.03}_{-0.07})$ & 24.2 & 3.2 (4.5) & $-24.82$ & 0.04 $(^{+0.16}_{-0.02})$ & 33.9 & 2.1 $(^{+15.1}_{-3.5})$ & 0.59 & 0.01 & $-693$ & $-390$ \\ 
 &      &                    &      &              & $-23.25$ & 0.22 $(^{+0.26}_{-0.72})$ & 8.6 & 0.9 $(^{+3.2}_{-0.6})$ & 0.69 & 0.01 &    &    \\ 
\\
{\bf Average}     & & & & & & & & & & &  $-4718$ & $-1039$  \\
{\bf Average/dof} & & & & & & & & & & & $-786$ &  $-208$  \\

\enddata
\tablecomments{ In the last two columns $\Delta\chi^2_d \equiv \chi^2_{2-deV}-\chi^2_{de~Vaucouleurs}$ and
$\Delta\chi^2_S\equiv\chi^2_{2-deV}-\chi^2_{Sersic}$.
Parameter uncertainties corresponding to the $1\sigma$ sky level
uncertainties are given in parentheses. As in Tables \ref{tab:devparams} and \ref{tab:serparams}, the absolute
magnitudes are {\sl not} corrected for extinction or k-dimming. These absolute magnitudes are total magnitudes,
which include extrapolation beyond $r=300$ \ho kpc.}

\label{tab:ddparams}
\end{deluxetable}

\begin{deluxetable}{lll}
\tablecaption{Abell 1651 \ser  Parameters}
\tablewidth{0pt}
\tablehead{
\colhead{} & \colhead{$r_e$} & \colhead{$n$} 
}
\startdata
Gonzalez et al. (2000)  & 59.6$\pm$1.1	& 4.3$\pm$0.2 \\
This work		& 65.0$^{+6.7}_{-5.6}$  & $4.26^{+0.17}_{-0.16}$ \\
\enddata

\label{tab:a1651}
\end{deluxetable}

\begin{deluxetable}{lllllllll}
\tablecaption{Average Absolute Magnitudes}
\tablewidth{0pt}
\tablehead{
\colhead{Model} & \colhead{$M_{Total}$} & \colhead{$\sigma_{M_{Total}}$} & \colhead{$M_{300}$} & \colhead{$\sigma_{M_{300}}$} &
 \colhead{$M_{50}$} & \colhead{$\sigma_{M_{50}}$} & \colhead{$M_{10}$} & \colhead{$\sigma_{M_{10}}$}    
}
\startdata
de Vaucouleurs & $-24.60$ & 0.56 & $-24.56$ & 0.53 & $-24.20$ & 0.39 & $-23.14$ & 0.30   \\
S\'ersic       & $-25.43$ & 0.83 & $-24.93$ & 0.50 & $-24.26$ & 0.36 & $-23.11$ & 0.28   \\
2-deV          & $-25.23$ & 0.59 & $-24.96$ & 0.47 & $-24.26$ & 0.35 & $-23.11$ & 0.29   \\
\enddata

\label{tab:global}
\tablecomments{The average absolute magnitudes and the dispersion in these magnitudes for our BCG sample. The column $M_{Total}$ corresponds to the total absolute magnitudes listed
in Tables \ref{tab:devparams} - \ref{tab:ddparams}. The columns $M_{300}$, $M_{50}$, and $M_{10}$ correspond to
the model magnitudes within circular metric apertures of radii 300, 50, and 10 \ho kpc, respectively. The above table does not include extinction or k-corrections, but both effects are small. Accounting for these factors decreases the dispersions by only 0.01 mag. The impact on the mean absolute magnitude can be approximated by assuming a mean extinction of 0.05 mag and a mean k-correction of 0.07 mag.
}
\end{deluxetable}

\begin{deluxetable}{lccccccccr}
\tablecaption{Impact of PSF on Derived \ddev Profile for Abell 0122 }
\tablewidth{0pt}
\tablehead{
\colhead{PSF} & \colhead{FWHM($\arcsec$)} & \colhead{$\mu_{e,1}$} & \colhead{$r_{e,1}$} &\colhead{$(b/a)_1$} & \colhead{$\mu_{e,2}$} & \colhead{$r_{e,2}$ (kpc)} & \colhead {$(b/a)_2$} & \colhead{$|\Delta\theta|$} &\colhead {$\Delta\chi^2$}
}
\startdata
Abell 0122 & 1.38 & 18.77 & 5.9 & 0.63 & 19.67 & 147.8 & 0.53 & 76.5 & -- \\
Abell 2376 & 1.33 & 18.83 & 7.4 & 0.67 & 19.74 & 157.9 & 0.52 & 72.6 & 51.7 \\
Abell 3727 & 1.37 & 18.81 & 6.6 & 0.63 & 19.71 & 152.6 & 0.52 & 75.3 & 82.4 \\
Abell S1096& 1.49 & 18.80 & 6.6 & 0.63 & 19.71 & 153.8 & 0.53 & 73.3 & -23.9 \\
\enddata

\tablecomments{ $\Delta\chi^2$ is the difference between the $\chi^2$ using this PSF and the $\chi^2$ using the correct PSF for the Abell 0122 field.
}
\label{tab:psfsensitivity}
\end{deluxetable}

\end{document}